\documentclass[nofootinbib,aps,11pt]{revtex4}
\usepackage[dvips]{graphicx}

\usepackage{amsmath}
\usepackage{bm}
\usepackage{times}
\usepackage{epsfig}
\usepackage{graphicx}
\usepackage{color}
\usepackage{cancel}
\usepackage{amssymb}
\usepackage{textcomp}

\newcommand{\Tr}{\text{Tr}}

\newcommand{\lsim}{\buildrel < \over {_\sim}}
\newcommand{\gsim}{\buildrel > \over {_\sim}}
\newcommand{\be}{\begin{equation}}
\newcommand{\ee}{\end{equation}}
\newcommand{\bea}{\begin{eqnarray}}
\newcommand{\eea}{\end{eqnarray}}
\newcommand{\ba}{\begin{array}}
\newcommand{\ea}{\end{array}}

\def\lsim{\mathrel{\raise.3ex\hbox{$<$\kern-.75em\lower1ex\hbox{$\sim$}}}}
\def\gsim{\mathrel{\raise.3ex\hbox{$>$\kern-.75em\lower1ex\hbox{$\sim$}}}}

\def\to{\rightarrow}

\def\beq{\begin{equation}}
\def\eeq{\end{equation}}
\def\be{\begin{equation}}
\def\ee{\end{equation}}
\def\bea{\begin{eqnarray}}
\def\eea{\end{eqnarray}}
\def\to{\rightarrow}

\def\beq{\begin{equation}}
\def\eeq{\end{equation}}
\newcounter{Lcount}

\begin{document}

\begin{flushright}
NPAC-08-20\\
MADPH-08-1517\\
IPMU-08-0091
\end{flushright}

\title{\Large Triplet Scalars and Dark Matter at the LHC}
\author{Pavel Fileviez P{\'e}rez$^{1}$, Hiren H. Patel$^{1}$, Michael. J. Ramsey-Musolf$^{1,2}$, Kai Wang$^{3,1}$}
\affiliation{
$^{1}$ University of Wisconsin-Madison, Department of Physics,
1150 University Avenue, Madison, WI 53706, USA
\\
$^{2}$ Kellogg Radiation Laboratory, California Institute of Technology,
Pasadena, CA 91125 USA
\\
$^{3}$ Institute for the Physics and Mathematics of the Universe,
University of Tokyo, Kashiwa, Chiba 277-8568, JAPAN}
\date{\today}
\pagestyle{plain}
\begin{abstract}
We investigate the predictions of a simple extension of the Standard Model
where the Higgs sector is composed of one $SU(2)_L$ doublet and one real triplet.
We discuss the general features of the model, including its vacuum structure,
theoretical and phenomenological constraints, and expectations for
Higgs collider studies. The model predicts the existence of a pair of light
charged scalars and, for vanishing triplet vacuum expectation value,
contains a cold dark matter candidate. When the latter possibility occurs,
the charged scalars are long-lived, leading to a prediction of distinctive
single charged track with missing transverse energy or double charged track
events at the LHC. The model predicts a significant excess of two-photon
events compared to SM expectations due to the presence of a light charged scalar.
\end{abstract}
\maketitle
\section{Introduction}
Uncovering the mechanism for electroweak symmetry-breaking (EWSB) is one
of the primary goals of the Large Hadron Collider (LHC). Despite the
tremendous successes of the Standard Model (SM), the scalar sector of
the theory that purports to be responsible for EWSB  has yet to be
confirmed experimentally. It is possible that the mechanism of EWSB
is more complicated than in the SM and that the low-energy scalar
sector contains more degrees of freedom than a single SU(2)$_L$ doublet.
A variety of extensions of the SM scalar sector have been proposed over
the years, and many of these introduce additional TeV-scale particles
in order to address other issues that cannot be resolved in the SM:
the gauge hierarchy problem, the abundance of matter in the universe
(both luminous and dark), gauge coupling unification, and the tiny
but non-vanishing neutrino masses.  In addition, the tension
between electroweak precision observables (EWPOs) that favor a
relatively light SM Higgs boson ($m_H = 84^{+33}_{-24}$ GeV~\cite{EWPO,Kile,LEP})
and the LEP II direct search lower bound $m_H\geq 114$ GeV~\cite{LEP2} point toward
the possibility of an augmented scalar sector with additional
light degrees of freedom.

The imminent operation of the LHC -- together with the recent establishment
of non-vanishing neutrino masses and heightened interest in the origin of
visible and dark matter -- make a detailed analysis of various scalar
sector extensions an important study. In this paper, we focus on the
possibility that the SM Higgs doublet is accompanied by a
light real triplet $\Sigma=(\Sigma^+,\Sigma^0,\Sigma^-)$ that transforms
as $(1,3,0)$ under SU(3$)_C\times$SU(2$)_L\times$U(1$)_Y$. This possibility
was first discussed by Ross and Veltman in Ref.~\cite{Veltman} and
subsequently by the authors of Refs.~\cite{triplet1,triplet2,triplet3,triplet4,triplet5,triplet6,Chivukula:2007gi}.
In Ref.~\cite{Cirelli:2005uq}, it was noted that the neutral component of $\Sigma$
could be a viable cold dark matter (CDM) candidate if it has no vacuum expectation
value. In that work, it was shown that the $\Sigma^0$ could saturate the
observed relic density, $\Omega_\mathrm{CDM}=0.1143\pm 0.0034$ \cite{Komatsu:2008hk}, if
$M_\Sigma \approx 2.5$ TeV. Since $\Omega_\mathrm{CDM}$ is reduced for
smaller $M_\Sigma$ due to the larger annihilation rate, a lighter triplet
would comprise one part of a multi-component dark matter scenario.

Recently, it was also observed in Refs.~\cite{GUT1,GUT2,GUT3} that in
several non-supersymmetric grand unified models that avoid rapid proton
decay and achieve coupling unification in agreement with experimental data,
a light real triplet emerges. In particular, as noted in Ref.~\cite{GUT1},
if the $SU(2)_L$ real triplet living in the adjoint representation $24_H$
of $SU(5)$ is light, it can help to achieve unification. From this standpoint, 
the model studied by Ross and Veltman in Ref.~\cite{Veltman} has a well defined 
UV completion, thereby providing extra motivation to study its phenomenological
aspects in detail.

In exploring the model's phenomenology, we will attempt to identify the main
features that distinguish it from other simple extensions of the SM scalar
sector, such as those with multiple SU(2$)_L$ doublets, $H^{'} \sim (1,2,1/2)$,
an extra real singlet, $S \sim (1,1,0)$, or a complex
triplet~\cite{Veltman} $\Delta\sim(1,3,1)$. In brief:
\begin{itemize}
\item[(i)] Models containing either a SM singlet or a second doublet can lead to
neutral scalar mass eigenstates that involve mixtures of the weak states.
The presence of this mixing can modify the tension between EWPO and direct
searches by allowing for lighter scalars to contribute to the renormalized SM
gauge boson propagators while reducing the Higgstrahlung production cross section
in $e^+e^-$ annihilation. Typically, the branching ratios for the decay
of the SM-like neutral mass eigenstate ($H_1^0$) are unchanged from
those of the SM Higgs, while the heavier neutral scalar ($H_2^0$) decays
can be different due to the presence of the \lq\lq Higgs splitting" decay mode:
$H_2^0 \to H_1^0 H_1^0$. Under some circumstances one has $M_{H_1} > 2 M_{H_2}$,
leading to a reduction in Br($H_1\to \mathrm{SM})$. In addition, models with
two Higgs doublets lead to an additional CP-odd scalar ($A^0$) and physical
charged Higgses ($H^\pm$) and one could have exotic Higgs properties
such as vanishing couplings to matter (\lq\lq fermiophobia").

In contrast, for models containing both an SU(2$)_L$ doublet and triplet, mixing
between neutral flavor states is generally suppressed due to constraints arising
from the $\rho$-parameter. Consequently, the effect on EWPO is typically less
pronounced than in the singlet or multiple doublet models, and the modification
of SM-like Higgs production is not sufficiently large to allow one to evade
the LEP II bounds. On the other hand, the $\Sigma^0$ can be stable, as noted above.
In this case, one can expect a relatively long-lived charged scalar, leading to
the possibility of distinctive charged track events at colliders. When the neutral
triplet-like scalar is not stable, its branching ratios can differ significantly
from those of the heavier neutral scalar in the singlet or two Higgs doublet
scenarios, due to differences in the couplings to gauge bosons.

\item[(ii)] The complex and real triplet scenarios lead to distinctive
features in both production and decay. For example, a complex triplet
(as in left-right symmetric models~\cite{LR}), $\Delta \sim (1,3,1)$,
couples to SM leptons leading to the Type-II see-saw mechanism~\cite{TypeII} for
neutrino masses. In this case one has the possibility of observing lepton number
violation through the decays $H^{++}\to e^+_i e^+_j$ and using the associated
production $H^{++} H^{-}$ one can distinguish this model easily~\cite{Tao}.
\end{itemize}

In what follows, we focus on the extension of the SM with a real triplet,
which we denote the \lq\lq $\Sigma$SM", and explore all features in detail.
The model predicts the existence of light charged Higgses that can be
considered as pseudo-Goldstone bosons. We find that in the $\Sigma$SM
the predictions for the decay of the SM-like Higgs into two photons
can differ substantially from the predictions in the Standard Model
due to contributions from the light charged scalar to the one-loop
decay amplitude. In the case when one assumes that the neutral
triplet-like Higgs has a vanishing vev and  is responsible for a
fraction of the cold dark matter density in the Universe, one expects
the charged scalars to be long-lived, leading to distinctive single or double
charged track plus $\not\!\! E_T$ events at the LHC. For non-vanishing
triplet vev, the two-photon decays of the triplet-like neutral scalar
can lead to a substantial rate for $\gamma\gamma \tau\nu$ and
$\gamma\gamma b{\bar b}$ final states in Drell-Yan production of
triplet-like scalar pairs. It may also be possible to discover
the $\Sigma$SM by searching for $b{\bar b}\tau+{\not\!\! E_T} $
events associated with the hadronic decays of the tau lepton.

This article is organized as follows: In section~\ref{sec:model} we
discuss the basic structure of the model that underlies these expectations,
including the various possibilities it provides for symmetry-breaking.
Section \ref{sec:pheno} gives the model's phenomenological
constraints, including those arising from EWPO and cosmology.
In sections \ref{sec:decay} and \ref{sec:production}, respectively,
we analyze the features of Higgs decays and production relevant
to both the LHC and Tevatron, including the dependence of these
features on the key parameters of the model as well as a detailed
study of the SM backgrounds. In particular, we discuss the prominent
signatures of the $\Sigma$SM  noted above. In the last section
we summarize the distinctive features of the model in comparison
with other scenarios for EWSB. A few technical details
appear in the Appendices.
\section{A Triplet Extension of the Standard Model}
\label{sec:model}
In this section we study the main properties of the triplet extension
of the Standard Model, wherein the scalar sector is composed of the SM Higgs,
$H \sim (1,2,1/2)$, and a real triplet, $\Sigma \sim (1,3,0)$. The Lagrangian of the
scalar sector is given by
\begin{equation}
\label{eq:lscalar}
\mathcal{L}_\mathrm{scalar} = (D_\mu H)^{\dagger} (D^\mu H) + \Tr (D_\mu \Sigma)^{\dagger}
(D^\mu \Sigma) - V(H,\Sigma) \,
\end{equation}
where $H^T=( \phi^+,\,\phi^0)$ is the SM Higgs and the real triplet can be written as
\begin{equation}
\Sigma = \frac{1}{2} \left( \begin{array} {cc}
\Sigma^0  &  \sqrt{2} \Sigma^+ \\
\sqrt{2} \Sigma^-  & - \Sigma^0
\end{array} \right) \,
\end{equation}
with $\Sigma^0$ being real, $\Sigma^+=(\Sigma^-)^*$ and
\bea
D_\mu\Sigma & = & \partial_\mu\Sigma +ig[{\tilde A}_\mu,\Sigma], \ \text{where} \ \qquad
{\tilde A}_\mu = \sum_{a=1}^3\, A_\mu^a T^a \ .
\eea
Here $A_\mu^a$ and $T^a$ are the gauge bosons and the generators of the group.
The most general renormalizable scalar potential is
\begin{eqnarray}
\label{scalar1}
V(H,\Sigma) & = & - \mu^2 \ H^\dagger H \ + \ \lambda_0  \ (H^\dagger H)^2 \ - \ M^2_{\Sigma} \ \Tr \ \Sigma^2
\ + \ \lambda_1 \ \Tr \ \Sigma^4 \ + \ \lambda_2 \ (\Tr \ \Sigma^2 )^2 \nonumber  \\
&& \qquad + \ \alpha \ ( H^\dagger H )\Tr \ \Sigma^2 \ + \ \beta \ H^\dagger \Sigma^2 H \
+ \ a_1 \ H^\dagger \Sigma H \ ,
\end{eqnarray}
where all parameters are real. Notice that Tr $\Sigma^{n}=0$, with $n$ odd.
We present a more compact form of the potential,
\bea
V(H,\Sigma) & = & - \mu^2 \ H^\dagger H \ + \ \lambda_0  \ \left( H^\dagger H \right)^2 \ - \frac{1}{2}\ M^2_{\Sigma} F \ + \ \frac{b_4}{4} F^2 \ + \ a_1 \ H^\dagger \Sigma H +\frac{a_2}{2} H^\dag H F \ ,
\label{eq:scalar1b}
\eea
where we have made the abbreviation $F\equiv \left(\Sigma^0\right)^2 + 2\Sigma^+\Sigma^-$, with
\be
b_4 = \lambda_2+\frac{\lambda_1}{2}, \qquad \text{and} \qquad a_2 = \alpha +\frac{\beta}{2} \ .
\ee
We emphasize that in the limit $a_1 \to 0$ (in the absence of the last term
in the potential in Eq.~(\ref{scalar1})) the scalar potential
of the theory possesses a global symmetry $O(4)_H \times O(3)_\Sigma$ and the
discrete symmetry $\Sigma \to - \Sigma$. These symmetries protect
the dimensionful parameter $a_1$, and the case of small $a_1$ corresponds
to a soft breaking of this symmetry.  We take advantage of the final term in the potential
in Eq.~(\ref{scalar1}) to establish the convention
that $a_1>0$ by absorbing the sign into the definition of $\Sigma$.
\subsection{Mass Spectrum and Vacuum Structure}
In general, the neutral components of both scalars, $H$ and $\Sigma$,
can have a non-zero vacuum expectation value. Defining
\bea
\label{eq:shifted}
H  & =&
\left(\begin{array}{c} \phi^+ \\ (v_0+h^0+i\xi^0) /\sqrt{2} \end{array}\right), \qquad \text{and} \ \qquad \
\Sigma  =  \frac{1}{2} \left( \begin{array}{cc}
x_0+\sigma^0  & \sqrt{2}\Sigma^+ \\
\sqrt{2}\Sigma^-  & - x_0-\sigma^0
\end{array} \right) \ ,
\eea
where $v_0$ and $x_0$ are the SM Higgs and triplet scalar vevs, respectively, we find that
the minimization conditions for the tree-level potential are
\begin{eqnarray}
\left( - \mu^2 +  \lambda_0  v_0^2  -  \frac{a_1  x_0 }{2} +   \frac{a_2  x_0^2 }{2} \right)  v_0 &=& 0 \ ,
\label{min1}
\\
- M_{\Sigma}^2  x_0  +  b_4 x_0^3  -  \frac{a_1  v_0^2}{4 } + \frac{a_2  v_0^2  x_0}{2} & = & 0 \ ,
\label{min2}
\end{eqnarray}
and
\begin{eqnarray}
\label{eq:curve}
b_4  >  \frac{1}{8 x_0^2} \left( - \frac{a_1 v_0^2}{x_0}  +  \frac{\left(- a_1 + 2 a_2 x_0 \right)^2}{2 \lambda_0} \right) \ , &&
\end{eqnarray}
where the last expression follows from the condition of a local minimum, i.e.
the determinant of the matrix containing the second derivatives must
be positive in each minimum. These conditions will, of course, require
modification when the full one-loop effective potential is considered.
For purposes of analyzing the basic phenomenological features of the model,
however, it suffices to consider the tree-level potential.

The minimization conditions of Eqs.~(\ref{min1}) and (\ref{min2}) allow for
four possible cases:\\
\indent\begin{tabular}{p{5cm}p{5cm}}
(1)\enspace $v_0\not=0$ and $x_0\not=0$ & (2) \enspace $v_0\not=0$ and $x_0=0$\\
(3)\enspace $v_0=0$ and $x_0\not=0$ & (4) \enspace $v_0=0$ and $x_0=0$
\end{tabular}\\
The last two possibilities are clearly not viable phenomenologically,
whereas either of the first two are, in principle, consistent with experiment.
The parameters in the potential must be chosen so that the global
minimum of the potential yields solutions (1)
and (2)\footnote{It is possible that the vacua with  $v_0\not=0$ are long-lived metastable minima\cite{Barger:2003rs,Kusenko:1996xt}, a possibility we do not consider here.}.
In addition, from Eq.~(\ref{min2}) we see that if $a_1\not=0$, solution (2)
is forbidden. Thus, a necessary (but not sufficient) condition
for a minimum with  $x_0=0$ but $v_0\not=0$ is that the model possesses 
the $O_{\Sigma}(3)$ global symmetry and $\Sigma \to - \Sigma$ symmetry.
The potential in Eq.~(\ref{scalar1}) is bounded from below
when $\lambda_0$ and $b_4$ are non-negative and when the
following relation holds for negative $a_2$:
\begin{equation}
a_2^2< 4 \lambda_0 \ b_4.
\label{bounded}
\end{equation}
In addition, before imposing the constraints coming from the mass spectrum,
the conditions $|\lambda_0| \leq 2 \sqrt{\pi}$, $|b_4| \leq 2 \sqrt{\pi}$,
and $|a_2| \leq 2 \sqrt{\pi}$ must be satisfied in order to keep perturbativity.
In what follows, we analyze the spectrum associated with different
phenomenologically viable vacua assuming each is the global minimum.
\subsubsection{\bf Mass Spectrum}
\noindent{\em Case (1a): $v_0\not=0$ and $x_0\not=0$ with $a_1\not=0$}\\
Upon electroweak symmetry breaking, the mass matrices of the neutral ($h^0$ and $\sigma^0$) and charged ($\phi^\pm$ and $\Sigma^\pm$) scalars, defined in Eq.~(\ref{eq:shifted}), are
\begin{equation}
\label{eq:massmtrx}
{\cal M}_{0}^2 = \left( \begin{array} {cc}
2 \lambda_0 v_0^2  &  - a_1 v_0 / 2 + a_2  v_0  x_0 \\
- a_1  v_0/ 2  +  a_2  v_0  x_0  & 2 b_4  x_0^2 + \frac{a_1  v_0^2}{4  x_0}
\end{array} \right), \enspace\text{and} \enspace
{\cal M}_{\pm}^2 = \left( \begin{array} {cc}
a_1 x_0  &  a_1 v_0 / 2  \\
a_1  v_0/ 2 & \frac{a_1 v_0^2}{4 x_0}
\end{array} \right) \ ,
\end{equation}
respectively, where the minimization conditions have been used to eliminate $\mu^2$ and $M_\Sigma^2$ in favor of the vacuum expectation values, $v_0$ and $x_0$.  The eigenvalues of these matrices are the tree-level masses of the physical scalars ($H_{1}$, $H_2$, $H^\pm$) of the theory, and are given by
\bea
\label{mneut1}
M_{H_1}^2 & = & \lambda_0 v_0^2 \left(1+\vert\csc 2\theta_0\vert\right) + \left( \frac{a_1 v_0^2}{8x_0} + b_4 x_0^2\right)\left(1-\vert\csc 2\theta_0\vert\right)\,, \\
M_{H_2}^2 & = & \lambda_0 v_0^2 \left(1-\vert\csc 2\theta_0\vert\right) + \left(\frac{a_1 v_0^2}{8x_0} + b_4 x_0^2\right)\left(1+\vert\csc 2\theta_0\vert\right), \enspace\text{and}\\
\label{eq:mpls1}M_{H^{\pm}}^2 &=&  a_1 x_0 \left( 1 + \frac{ v_0^2}{4 x_0^2} \right) \,,
\eea
where $\theta_0$ is a mixing angle defined below, in Eq~(\ref{eq:neutralmix}) and csc stands for cosecant.
The mass parameters of the $\xi^0$ field and the second eigenvalue of $\mathcal{M}_\pm^2$ are vanishing, 
and are associated with the would-be Goldstone bosons, $G^0$ and $G^\pm$ respectively. The physical mass 
eigenstates and the unphysical electroweak eigenstates are related by rotations through two new mixing 
angles -- one for the neutral scalars, $\theta_0$, and the other for charged scalars $\theta_+$:
\begin{eqnarray}
\label{eq:neutralmix}
\left( \begin{array}{c} H_1 \\ H_2\end{array} \right) & = & \left(\begin{array}{ccc} \cos \theta_0 & \sin \theta_0 \\ - \sin \theta_0 & \cos \theta_0\end{array} \right) \left( \begin{array}{c} h^0 \\ \sigma^0 \end{array} \right) \ , \hspace{10mm} G^0=\xi^0\,,\\
& & \nonumber \\
\label{eq:chargemix}
\left( \begin{array}{c} H^\pm \\ G^\pm \end{array} \right) & = & \left(\begin{array}{cc} -\sin \theta_\pm & \cos \theta_\pm  \\ \cos\theta_\pm & \sin \theta_\pm  \end{array} \right) \left( \begin{array}{c} \phi^\pm \\ \Sigma^\pm \end{array} \right) \ .
\end{eqnarray}
In terms of parameters in the Lagrangian, the mixing angles are
\begin{equation}
\label{eq:thetazero}
\tan 2 \theta_0 =  \frac{4 v_0 x_0(-a_1 + 2 x_0 a_2)}{8 \lambda_0 v_0^2 x_0 - 8
b_4 x_0^3 - a_1 v_0^2},\hspace{5mm}\text{and}\hspace{5mm}\tan 2 \theta_+ =  \frac{4 v_0 x_0}{4 x_0^2-v_0^2}\,.
\end{equation}
The neutral mixing angle $\theta_0$ can, in turn, be expressed in terms of the physical masses:
\begin{equation}
\label{eq:rdef}
\tan 2 \theta_0 =  \frac{2x_0}{v_0}r,\hspace{5mm}\text{with}\hspace{5mm}r  \equiv
\frac{a_2  v_0^2 - 2  M_{H^+}^2}
{M_{H_1}^2  +  M_{H_2}^2  - 2  M_{H^+}^2  - 4 b_4 x^2_0} \ .
\end{equation}
We note that the mass-squared of the charged Higgs, Eq.~(\ref{eq:mpls1}),
is linearly proportional to $a_1$. Since, as we previously mentioned that,
in the limit $a_1 \to 0$, the theory enjoys a global $O(3)_{\Sigma}$
symmetry, we identify these charged scalars, $H^\pm$, as the associated
pseudo-Goldstone bosons for small $a_1$.

We will elaborate in more detail in Section \ref{sec:pheno} that constraints
coming from measurements on the $\rho$-parameter place an upper bound on the
triplet vev, $x_0$, which we take to be $(2x_0/v_0)^2\lsim0.001$. Since
the neutral mixing angle, $\theta_0$, is proportional to $x_0/v_0$, it
remains small throughout the parameter space, except when $M_{H^+}^2 \approx (M_{H_1}^2  +  M_{H_2}^2)/2$.  For this reason,
we refer to $H_1$ as the SM-like scalar and $H_2$ as the $\Sigma$-like scalar.  Using the condition in Eq.~(\ref{bounded})
and the approximation that $M_{H_1}^2 \approx 2 \lambda_0 v_0^2$ we find that $b_4 > 0$. Therefore,
$0 < \lambda_0\ , b_4 < 2 \sqrt{\pi}$.
\\

\noindent{\em Case 1b): $v_0\not=0$ and $x_0\not=0$ with $a_1=0$}

After EWSB that leads to $v_0 \not=0$, the $\Sigma$SM retains an $O(3)_{\Sigma}$ global symmetry
as well the discrete $\Sigma\to -\Sigma$. The breaking of the global $O(3)_{\Sigma}$ implies
the existence of massless Goldstone bosons\footnote{These are the same Goldstone bosons
of the model proposed by Georgi-Glashow in 1972~\cite{GG72}.} -- in this case, the $\Sigma^\pm$ --
in addition to the SM would be Goldstone bosons. From Eq.~(\ref{eq:massmtrx}) and the vanishing
of $\mathcal{M}^2_\pm$ with $a_1$, we see the appearance of this second massless
mode explicitly. The presence of these massless charged scalars with unsuppressed
gauge coupling to the $Z^0$ is precluded by LEP studies, so that this case is ruled out by experiment.
Given these considerations, we do not consider this case further, and we will avoid any choice
of the parameters in the potential implying a global minimum for $v_0\not=0$ and $x_0\not=0$ with $a_1=0$.
When $a_1=0$ and $x_0\not=0$, the charged scalars are massless at tree-level
as indicated by Eqs.~(\ref{eq:massmtrx}) and (\ref{eq:mpls1}).

\noindent{\em Case (2):  $v_0\not=0$ and $x_0=0$}

For this scenario, wherein $a_1$ and $x_0$ both vanish, $H$ and $\Sigma$ do not mix
and the tree-level masses are given by
\bea
\label{eq:dmmass}
M_{H_1}^2 & = & 2 \lambda_0 v_0^2 \ ,
\eea
and
\bea
\label{eq:dmmass2}
M_{H_2}^2=M_{H^\pm}^2 & = & -M_\Sigma^2 + \frac{a_2 v_0^2}{2}\equiv M_0^2 \ .
\eea
Radiative corrections break the degeneracy between the charged
and neutral components of the triplet. The mass splitting has
been computed in Ref.~\cite{Cirelli:2005uq}
\be
\label{eq:deltam}
\Delta M\equiv M_{H^\pm}-M_{H_2} = \frac{\alpha M_0}{4\pi s_W^2}\, \left[ f \left(\frac{M_W}{M_0}\right)-c_W^2 f\left(\frac{M_Z}{M_0}\right)\right] \ ,
\ee
where $s_W$ ($c_W$) gives the sine (cosine) of the weak mixing angle,
\bea
f(y) = -\frac{y}{4}\left[2y^3\ln y -ky+(y^2-4)^{3/2}\ln A\right],\qquad\text{with}\qquad A=\frac{1}{2}(y^2-2-y\sqrt{y^2-4})\,,
\eea
and $k$ contains the U.V. regulator. Note that when the tree-level
relation $c_W M_Z= M_W$ is used, the dependence of the mass
splitting on $k$ vanishes. The resulting value for the splitting
is
\be
\label{eq:masssplit}
\Delta M=(166\pm 1)\ \ \mathrm{MeV}
\ee
in the limit $M_0 \gg M_{W}$.
\subsubsection{\bf Vacuum Structure}
Having identified the four possibilities for symmetry breaking and
the corresponding scalar mass spectrum for those that remain
phenomenologically viable, we discuss in Appendix \ref{app:vac} the conditions under which the specified values of the doublet and triplet vevs yield the absolute minimum vacuum energy (we always require that specified vevs correspond at least to a local minimum). These considerations will place restrictions on the remaining independent model parameters for the two phenomenologically viable cases:
\begin{itemize}
\item[(1)] For this case, for which both vevs are non-vanishing, we eliminate $\mu^2$ and $M_\Sigma^2$ as independent parameters in favor of $v_0$, $x_0$ and the remaining four independent parameters: $\lambda_0$, $b_4$, $a_1$, and $a_2$. In the 
discussion of the low-energy phenomenology, we will trade three of the latter in terms of the physical masses, choosing as the six independent parameters: $M_{H_1}$, $M_{H_2}$, $M_{H^+}$, $v_0$, $x_0$, and $a_2$ with $v_0=246$ GeV.
\item[(2)] In this scenario with vanishing triplet vev and corresponding to triplet dark matter, we begin with five independent parameters since $a_1$ must vanish. Noting that $M_{H_2}=M_{H^+}$ at tree-level, we choose
$M_{H_1}$, $M_{H_2}$, $v_0$, $a_2$, and $b_4$ as independent parameters.
\end{itemize}
When discussing the low-energy phenomenology, we will give the dependence of branching
ratios and collider production rates on $M_{H_1}$, $M_{H_2}$, $M_{H^+}$, $x_0$,
and $a_2$ without imposing the requirement of absolute vacuum energy minimum.
It is possible that the chosen minimum is not the absolute minimum but
rather a long-lived metastable minimum~\cite{Barger:2003rs,Kusenko:1996xt}.
Requiring that the lifetime of the metastable vacuum is much larger than
the age of the universe will lead to restrictions on the model parameters,
but these restrictions may be less severe than those following from the
requirement that the chosen vacuum is the absolute minimum. In the case of
the minimal supersymmetric Standard Model (MSSM) for example, it has been
shown in Ref.~\cite{Kusenko:1996jn} that the conditions on the third generation
triscalar couplings that follow from metastability of the electroweak minimum
with respect to a charge and color breaking minimum are considerably less
restrictive than those implied by taking the electroweak vacuum to be the
absolute minimum. A detailed analysis of the metastability conditions for
the $\Sigma$SM involves a substantial numerical investigation, which we defer
to future work. Instead, we outline in Appendix \ref{app:vac} the conditions
that are likely to be sufficient but not necessary for the universe
to have evolved into the specified vacuum.
\subsection{Interactions: Main Features}
The full set of interactions involving $H$, $\Sigma$, and 
gauge bosons follow from Eqs.~(\ref{eq:lscalar}-\ref{eq:scalar1b}) and
the mixing matrices in Eqs.~(\ref{eq:neutralmix}) and (\ref{eq:chargemix}).
The Feynman rules relevant to our analysis of the production and decay
phenomenology appear in the Appendix. Here we highlight a few key
features of these interactions and their implications for phenomenology.

\begin{itemize}

\item {\textit{ Higgs-Higgs Interactions}}: The terms in $V(H,\Sigma)$ proportional
to $a_1$ and $a_2$ provide for so-called \lq\lq Higgs splitting" decay modes
such as $H_2 \to H_1 H_1$ when kinematically allowed. Note that the amplitude for the Higgs splitting
decay of the neutral triplet-like scalar, $H_2$, is proportional to $x_0$ and is 
thus suppressed.

\item {\textit{ Gauge-Higgs Interactions}}: As usual, one has couplings of the type $\Sigma \Sigma V$
and $\Sigma \Sigma V V$ where $V=\gamma, Z, W^\pm$. The former are responsible for the
dominant production mode of the $H_2$ and $H^\pm$ through
the $q{\bar q}^\prime \to V^\ast \to H H$ pair production process. Both couplings
also contribute to the weak vector boson fusion (VBF) production process.
Couplings of the type $H VV^\prime$ where $H$ denotes $H_2$ or $H^\pm$ will be suppressed
either by $x_0$ or the small mixing between the SM-like and triplet-like scalars.
For this reason, associated production of a single triplet-like scalar, 
$H_2^0 Z$, $H_2^0 W$, and $H^{\pm} W^{\mp}$, will be strongly suppressed compared 
to the corresponding production of a SM-like scalar.

From the standpoint of decay profiles, the $x_0$ (or mixing factor)
suppression is generally not relevant, since it cancels from branching ratios. However, an exception occurs
in the case of the singly charged scalar, $H^\pm$, which has three relevant  couplings involving gauge bosons:  $H^{\pm} Z W^{\mp}$,
$H^{\pm} W^{\mp} H_1$, and  $H^{\pm} W^{\mp} H_2$. The first two couplings are proportional to $x_0$, while the latter contains a
component that is free from this suppression factor and that is generated by the underlying $\Sigma^\pm\Sigma^0 W^\mp$ interaction. Given the small mass splitting, Eq.~(\ref{eq:masssplit}), this interaction allows for the decay $H^\pm\to H_2\pi^\pm$ that occurs via the emission of a virtual $W^\pm$. In the limit of tiny $x_0$, this decay channel becomes the dominant one. In the case of the extra neutral Higgs, $H_2$, one
finds that there are two relevant couplings to gauge bosons $H_2 Z Z$
and $H_2 W^{\pm} W^{\mp}$,  both of which proportional to $x_0$. As we discuss below, these couplings contain distinct dependences on the quantity $r$ defined in Eq.~(\ref{eq:rdef}). In particular, the $H_2 W^{\pm} W^{\mp}$ vertex is
\be
\label{eq:H2WW}
H_2 W^\pm W^\mp\ : \quad i g^2 (2-r) x_0 g^{\mu\nu}
\ee
while the $H_2 ZZ$ and $H_2 f{\bar f}$ couplings are all proportional to $x_0 r$ (see below) since they occur only in the presence of $h^0$-$\Sigma^0$ mixing. The $r$-independent term in
Eq.~(\ref{eq:H2WW}) is generated by the $\Sigma^0\Sigma^0 W^+ W^-$ term in the Lagrangian after the $\Sigma^0$ obtains a vev. In contrast, there is no $\Sigma^0\Sigma^0 ZZ$ term or coupling of the $\Sigma^0$ to matter fields in the Lagrangian, so the $H_2 ZZ$ and $H_2 f{\bar f}$ vertices must be proportional to the mixing parameter $r$. As we discuss below, one may in principle exploit these different dependences on $r$ to study the $a_2$-dependence of various triplet-like scalar branching ratios.

\item {\textit{ Yukawa Interactions}}: When the mixing angles
are non-zero, both the SM-like and triplet-like scalars couple to
fermions through Yukawa interactions. The relevant part
of the Lagrangian describing interactions between the
physical scalars and the SM fermions is
\begin{eqnarray}
\mathcal{L}_\text{Yuk}&=&
\frac{m_f}{v_0} \cos \theta_0 \ \overline{f} f  H_1 \ - \
\frac{m_f}{v_0} \sin \theta_0 \ \overline{f}  f  H_2 \ - \ \nonumber
\\
&& - \frac{\sqrt{2}}{v_0} \sin \theta_{+} \ \bar{u} \left( - m_u \ V_{CKM} \ P_L \ + \ V_{CKM} \ m_d  \ P_R \right)
d \ H^{+} \ + \ \mathrm{h.c.}\ \ \ ,
\end{eqnarray}
where $f$ stands for any charged SM fermion and $V_{CKM}$ is the 
Cabibbo-Kobayaski-Maskawa matrix. 
Since $\theta_+ \sim x_0/v_0$, $\theta_0 \sim x_0/v_0$ and $x_0 \ll v_0$ 
the Yukawa couplings of $H_2$ and $H^{\pm}$ are always suppressed compared 
to those of the doublet-like neutral scalar. As discussed above, this suppression 
will not affect the $H_2$-decay branching ratios but does govern those 
of the $H^\pm$ which can decay to $H_2 \pi^\pm$--even for zero $x_0$.
\end{itemize}
We emphasize that the presence of gauge interactions involving the $\Sigma$ implies that the $H_2$ branching ratios are generally
different from those in other extended Higgs sector models that lead to a second, CP-even neutral scalar. For example, in
extensions involving a single real scalar singlet, $S$, the $H_2$ and $H_1$ branching ratios will be identical
when $M_{H_2} < 2 M_{H_1}$ since the $H_2$ can decay only due to  $S$-$h^0$ mixing. Modifications only occur when
the Higgs splitting mode $H_2\to H_1 H_1$ becomes kinematically allowed. In the $\Sigma$SM, on the other hand,
the $H_2$ coupling to $ZZ$ and $f{\bar f}$ can only occur at tree-level through $\Sigma^0$-$h^0$ mixing, while the
existence of its coupling to $W^+W^-$ does not require such mixing. Below the $WW$ threshold, this difference
will affect Br($H_2\to\gamma\gamma)$ which is dominated by $W$-boson loops, while above the $WW$ threshold, it will 
imply a difference between Br($H_2\to WW$) and Br($H_1\to WW$), even in the absence of a kinematically allowed Higgs splitting mode.
\subsection{Phenomenological Constraints}
\label{sec:pheno}
Electroweak precision observables (EWPO) and direct searches place important constraints
on the parameters of the model. Here we review the phenomenological constraints that
have the most significant impact on the prospects for discovering the $\Sigma$SM and
distinguishing it from other possibilities.
\begin{itemize}
\item {\textit {The $\rho$ parameter.}}
In this theory $\Sigma^0$ does not contribute to the $Z$ mass, since there is
no $(\Sigma^0)^2 Z^2$ interaction. It does, however contribute to $M_W$ through
a $(\Sigma^0)^2 W^+W^-$ interaction. Consequently, the gauge boson masses are
given at tree-level by
\begin{eqnarray}
M_W^2 &=& \frac{ g_2^2}{4} \left( v_0^2 + 4 x_0^2 \right), \qquad\text{and}\qquad
M_Z^2= \frac{g_1^2 + g_2^2}{4} v_0^2, \ \ \
\end{eqnarray}
leading to a well-known tree-level correction to the $\rho$-parameter:
\be
\rho = \frac{M_W^2}{M_Z^2\cos^2\hat\theta_W \hat\rho} = 1+\delta\rho,
\ee
where
\be
\cos^2\hat\theta_W = \frac{{\hat g}_2^2}{{\hat g}_2^2+{\hat g}_1^2},
\ee
gives the weak mixing angle in the $\overline{\mathrm{MS}}$ scheme
(indicated by the hatted quantities), $\hat\rho$ gives the effect of SM electroweak
radiative corrections, $\delta\rho$ denotes contributions from new physics.
In the present case, we have
\be
\delta\rho = \left(\frac{2x_0}{v_0}\right)^2\ \ \ .
\ee
From a global fit to EWPO one obtains the $1\sigma$ result
\be
\label{eq:rho1}
\delta\rho=0.0002^{+0.0007}_{-0.0004}\ \ \ .
\ee
Consequently, in what follows we will adopt the bound
\be
\left(\frac{2x_0}{v_0}\right)^2 \lsim 0.001 \ , \qquad \text{or} \qquad
x_0 \lsim 4\ \ \mathrm{GeV}\ .
\label{eq:rho3}
\ee
The bound in Eq.~(\ref{eq:rho3}) could be relaxed by requiring a higher level
of confidence, but the magnitude would not change by more than a factor of two.
Such a change would be inconsequential for the phenomenology
of the $\Sigma$SM, so we will retain the bound of Eq.~(\ref{eq:rho3}).
\item {\textit{ Corrections to the $W$ and $Z$ boson propagators.}}
Because the $\Sigma$ couples to electroweak gauge bosons, it will generate one
loop contributions to the corresponding propagators. These contributions have been
studied extensively in Refs.~\cite{triplet3,triplet4,triplet5,triplet6}.
In light of the $\rho$-parameter constraints on $x_0$ it is instructive to consider
these effects in the limit of vanishing mixing angle. As discussed above, this
limit can arise when either: $a_1$ and $x_0$ both vanish, or $a_1$ vanishes but not $x_0$.
When $a_1$ and $x_0$ both differ from zero, we may consider this limit as the
first term of an expansion in the small mixing angles. To that end, we will
consider the combinations of the gauge boson propagators
that appear in the oblique parameters $S$, $T$, and $U$. To zeroth order
in the mixing angles, $\theta_{0,+}$, the triplet contribution to $S$ vanishes
since $Y(\Sigma)=0$ \cite{triplet3}. The effects of $\Sigma$ on $S$ can only arise
through mixing with $H$, which carries unit hypercharge. At lowest order in gauge
interactions and zeroth order in mixing angles, $\theta_{0,+}$, the triplet contribution to
the $T$ parameter is small since it is protected by the custodial SU(2$)_L$ symmetry.
In this limit, the tree-level relation between the masses $M_{H_2}$
and $M_{H^\pm}$ is given by
\be
\label{eq:deltamsq}
\Delta M^2\equiv M_{H^\pm}^2-M_{H_2}^2\Bigr\vert_\mathrm{tree} =
\biggl\{
\begin{array}{cc}
a_1x_0-2b_4 x_0^2\ , & a_1\not=0,\, x_0\not=0,\\
0\ , & a_1=0=x_0\ \ \ .
\end{array}
\ee
The $T$ parameter is given by
\be
{\hat\alpha} T = \frac{1}{M_W^2}\left[ {\hat c}^2\left({\hat\Pi}_{ZZ}(0) +\frac{2{\hat s}}{\hat c}{\hat\Pi}_{Z\gamma}(0)\right)
-{\hat\Pi}_{WW}(0)\right]\,.
\ee
We find that in the limit of zero mixing,
${\hat\Pi}_{ZZ}(0)=0={\hat\Pi}_{Z\gamma}(0)$, while
\bea
\nonumber
{\hat\Pi}_{WW}(0)&=& -\frac{g^2_2}{16\pi^2}\left[\frac{1}{2}\left(M_{H^\pm}^2+M_{H_2}^2\right)-\frac{M_{H^\pm}^2 M_{H_2}^2}{M_{H_2}^2-M_{H^\pm}^2}\, \ln \frac{M_{H_2}^2}{M_{H^\pm}^2}\right]\\
&\approx& \frac{g^2_2}{24\pi^2}\left(M_{H^\pm}-M_{H_2}\right)^2\,,
\eea
where we have neglected terms of $\mathcal{O}(x_0/v_0)^2$.
From the bound in Eq.~(\ref{eq:rho3}) and the expression in Eq.~(\ref{eq:deltamsq})
we observe that  $|\Delta M^2|/M_W^2 <<1$. Using the relation
${\hat g}_2^2=4\pi{\hat\alpha}/{\hat s}^2$ we obtain
\be
T_{\Sigma} \approx  -\frac{1}{6\pi{\hat s}^2}\, \frac{(\Delta M)^2}{M_W^2} \ \ \ .
\ee
A global fit to all EWPO gives~\cite{Kile}
\be
T - T_{SM} = -0.111 \pm 0.109,
\ee
or
\be
-0.220 \leq T_\Sigma < -0.002,
\ee
at 68 \% confidence. The corresponding range for  the mass splitting is
\be
\label{eq:ewpodeltam}
0.009 \ M_W^2 \leq (\Delta M)^2 \leq 0.96 \ M_W^2\ \ \ .
\ee
The constraints on $\Delta M$ that follow from the $\rho$-parameter are
clearly consistent with this result. One-loop gauge boson contributions to
$\Delta M$ are much smaller than $M_W$ and do not affect our general conclusions\footnote{One should not interpret the 68\% C.L. lower bound in Eq.~(\ref{eq:ewpodeltam}) as implying a minimum mass splitting; the $2\sigma$ range, for example, is consistent with $\Delta M^2=0$.}. It is possible that 
the mixing angle $\theta_0$ is not small when $M_{H_1}^2+M_{H_2}^2\approx 2 M_{H^\pm}^2$ 
[see Eq.~(\ref{eq:rdef})]. 
This scenario could lead to substantial effects on the gauge boson propagators and may help alleviate the tension 
between EWPO that favor a light SM-like Higgs and the lower bound from direct searches. 
We will explore this possibility more extensively in a subsequent study and concentrate 
in this work on the small mixing scenario. See Ref.~\cite{Chen} for a recent study of 
these constraints.
\item {\textit{Collider Constraints.}}
LEP searches for both charged and neutral scalars place severe constraints on the possible existence
of light scalars. The neutral scalar Higgs $H_1$ is SM-like, and one has to impose the lower
bound from LEP2, $M_{H_1} > 114$ GeV. In the case of the singly charged Higgses, $H^{\pm}$,
one should assume a conservative lower bound $M_{H^{\pm}} \geq 100$ GeV due to the absence of non-SM 
events at LEP~\cite{LEP}. Since $M_{H_2} \approx M_{H^{\pm}}$ one has to use the same bound for the extra neutral scalar Higgs.
\item {\textit{Big Bang Nucleosynthesis.}}
In principle, considerations of primordial nucleosynthesis could have
important implications for the $\Sigma$SM. In particular, it has been pointed
out in Ref.~\cite{BBN} that the existence of a charged scalar with
lifetime $\tau >  10^{3}$ s can reduce the relative abundance of $^{6}$Li
produced during big bang nucleosynthesis, thereby exacerbating the
present tension with the $^2$H and $^4$He abundances and the value of the
baryon asymmetry derived from the cosmic microwave background. This bound is
irrelevant for the $\Sigma$SM, however, since  the decay $\Sigma^\pm\to H_2 W^{\pm\ast}\to H_2\pi^\pm$
is very fast (see Fig.~\ref{fig:hpmdecaylength} below).
\end{itemize}
\section{Properties of the Higgs Decays}
\label{sec:decay}
As discussed above, there are four physical scalars
in this theory: two neutral scalars $H_1$ and $H_2$ (SM-like and
triplet-like, respectively), and two singly charged
scalars $H^{\pm}$ with small couplings to fermions. In this section
we discuss the main features of the Higgs decays in all possible
scenarios.
\subsection{Cold Dark Matter and Higgs Decays}
In the case when the real triplet does not acquire a vev, the neutral component $\Sigma^0$ can
be a viable cold dark matter candidate. We previously mentioned that, in this case, the scalar
potential has a global $O(3)_\Sigma$ symmetry and a $\Sigma \to - \Sigma$, discrete symmetry. In Ref.~\cite{Cirelli:2005uq}
this CDM candidate has been studied in detail. Under the assumption that this candidate is responsible for
the CDM relic density in the Universe, the mass should be $M_{\Sigma} \approx 2.5$ TeV.  However, as we will show in
the next section the production cross section is very small in this case. In this scenario the main
decay channel of the singly charged Higgs is $H^+ \to H_2 \pi^+$ due to the small mass splitting coming from
radiative corrections. In order to test this scenario at the LHC, we must 
assume that $M_\Sigma\ll 2.5$ TeV so that the $\Sigma^0$ is only one component of the CDM density. 
In this case the pair production and weak vector-boson fusion cross sections
for $H^+ H_2$ and $H^+H^-$ are large enough to generate observable effect. Since the
$H_2\equiv\Sigma^0$ is stable one should only expect to see missing
energy and a charged track. In Ref.~\cite{Cirelli:2005uq}
the authors pointed out that if the mass of $H_2$ is
approximately $500$ GeV, its relic density makes up about
$10 \%$ of the total DM density. We will restrict our attention to the scenarios where $H_2$ is light
in order think about the possibility to test the model at the LHC.

\begin{figure}[ht]
\includegraphics[scale=1,width=8cm]{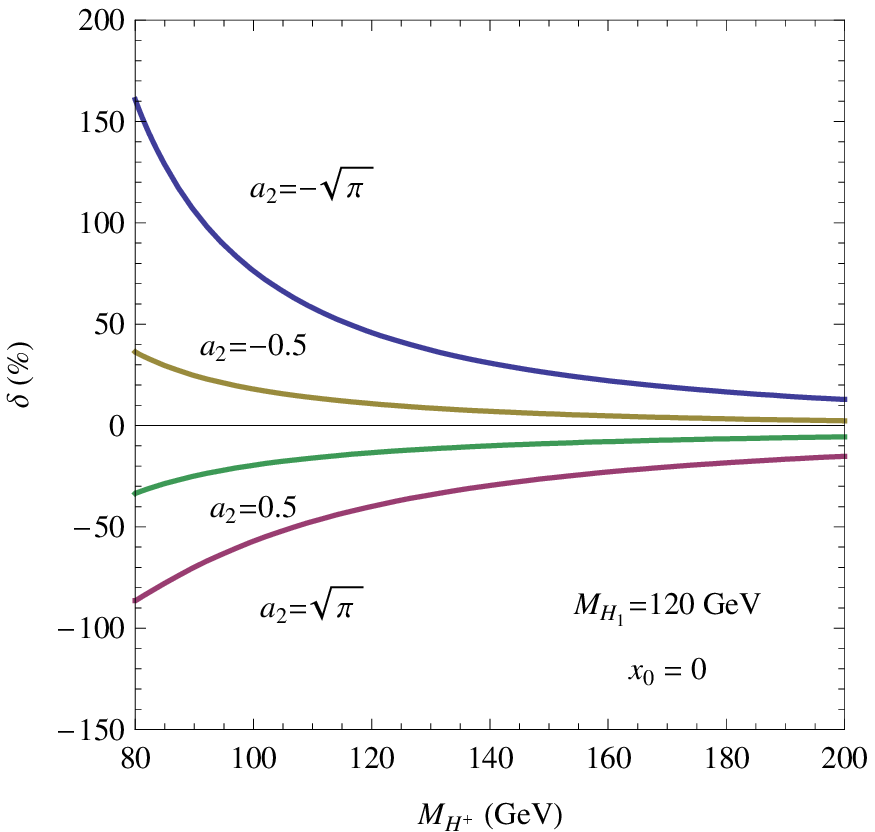}
\includegraphics[scale=1,width=7.7cm]{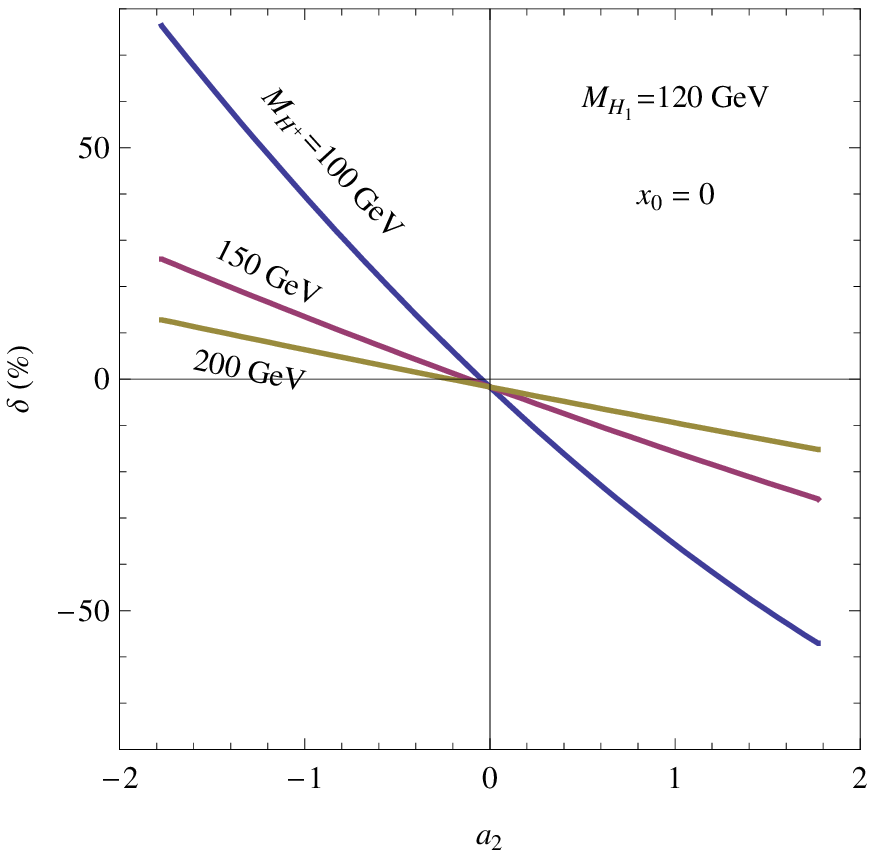}
\caption{Predictions for $\delta$, as defined in Eq.~(\ref{delta}),
in the case of $x_0=0$ and $M_{H_1}=120$ GeV.  Left panel shows the $\delta$ dependence on $M_{H^+}$. 
Different curves correspond to different values of $a_2$.  Right panel shows the $\delta$ dependence on $a_2$, with 
different curves corresponding to different charged Higgs masses, $M_{H^+}$.}
\label{H1ggx00}
\end{figure}

The existence of the charged scalars in $\Sigma$SM can modify
predictions for the decay of the SM-like Higgs, $H_1$, into
two photons since, in general, the $a_2$ parameter can be large.
This effect arises from the quartic $H^\dag H F^2$ term in the potential,
proportional to $a_2$, that generates a $h^0 \Sigma^+\Sigma^-\sim H_1 H^+ H^-$
coupling after EWSB. Note that this interaction is not suppressed by 
the triplet vev (see Appendix C for the Feynman rules). For a sufficiently light charged Higgs, $H^+$,
and large $|a_2|$, the charged scalar loop contributions to
the $H_1\to\gamma\gamma$ amplitude can yield non-negligible changes in Br($H_1\to\gamma\gamma$).
In order to analyze the impact of the charged Higgs in this mode,
we define the relative change in the $H_1 \to \gamma \gamma$ decay partial width by
\begin{equation}
\delta= \frac{\Gamma^{\Sigma} (H_1 \to \gamma \gamma) \ - \ \Gamma^\text{SM} (H_1 \to \gamma \gamma)}{ \Gamma^\text{SM} (H_1 \to \gamma \gamma)}\,,
\label{delta}
\end{equation}
where $\Gamma^\Sigma (H_1 \to \gamma \gamma)$ and $\Gamma^\text{SM} (H_1 \to \gamma \gamma)$ are
the decay widths with and without the contribution of the charged Higgs, respectively.  In Fig.~\ref{H1ggx00}
we show $\delta$ for $x_0=0$ and different values of the $a_2$ parameter and charged Higgs mass.
Notice that predictions for the decays into two photons can be modified appreciably
when the charged Higgs mass is below $200$ GeV. When the $a_2$ parameter is negative we
find a large enhancement in the decay width. Since, when $x_0=0$ where the
DM candidate, $H_2$, and the charged Higgs, $H^{+}$, are approxiately degenerate,
we expect large modifications of the decay mode $H_1 \to \gamma \gamma$ only when $H_2$
is responsible for a fraction of the Dark Matter density in the Universe.
%
\subsection{SM-Like Higgs Boson Decays: General Case of ${\bm x_0\not=0}$}
Since the mixing between the SM Higgs and the real triplet is typically small,
the scalar $H_1$ is SM-like.  The decays of $H_1$ are similar to the decays of the SM Higgs except for
the decays into two photons. As we have discussed before, the presence of the charged Higgs can dramatically modify the decay width for this channel. Since this channel is important for the discovery of the scalars at the LHC we discuss the predictions
here in detail. The expected accuracy for the branching ratio at the LHC for this 
channel is about $20\%$~\cite{EWPO}.

\begin{figure}[h]
\includegraphics[scale=1,width=8cm]{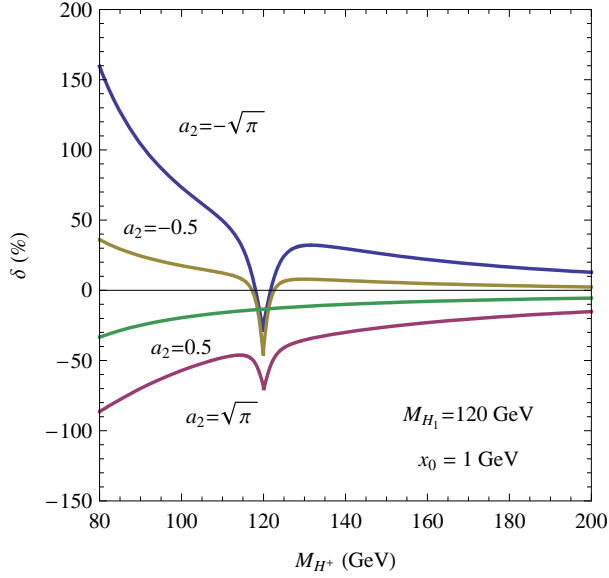}
\caption{Values for $\delta$ in percent, as defined in Eq.~(\ref{delta}), when $x_0=1$ GeV and $M_{H_1}=120$ GeV.  }
\label{H1ggx01}
\end{figure}

In Fig.~\ref{H1ggx01} we show the values for the difference between the
predictions in the SM and in our model for $H_1 \to \gamma \gamma$
when $x_0=1$ GeV and $M_{H_1}=120$ GeV. When $M_{H^+}\sim 120$ GeV,  $\delta$ is
small since the mixing angle is large and in this case the coupling between $H_1$
and $H^{\pm}$ is suppressed when $a_2$ is negative. Apart from this particular 
region of parameter space, we expect a large modification of the decay width of the SM-like Higgs
decay into two photons when $x_0\not=0$. More generally, for light $H^\pm$, the relative change in the
$\Gamma(H_1\to\gamma\gamma)$ can be larger in magnitude than the expected LHC 
precision for this channel\cite{EWPO}, allowing one to use this channel 
to gain indication of the sign of the $a_2$ coupling over a limited range 
of the parameter space. As we discuss below, one may in principle determine 
$M_{H^+}$ by studying its branching ratios. Looking further to the future, a more precise study of
Br($H_1\to\gamma\gamma$) at an $e^+e^-$ collider could be carried out~\cite{ILC}.
%
\subsection{Charged Higgs Boson Decays}
\begin{figure}[h]
\includegraphics[scale=1,width=0.45\textwidth]{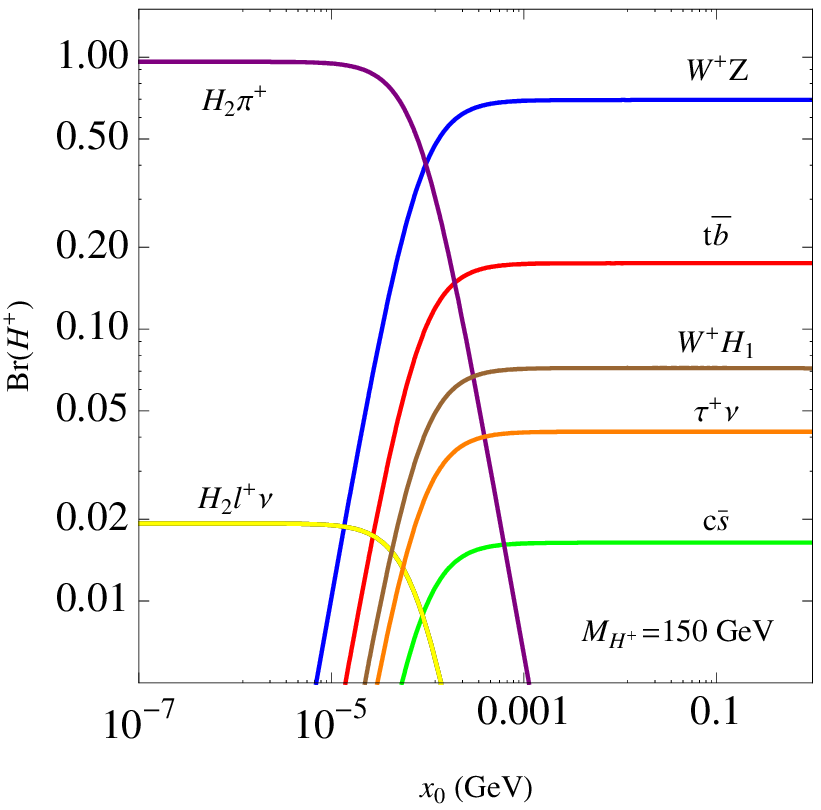}\hspace{0.03\textwidth}
\includegraphics[scale=1,width=0.45\textwidth]{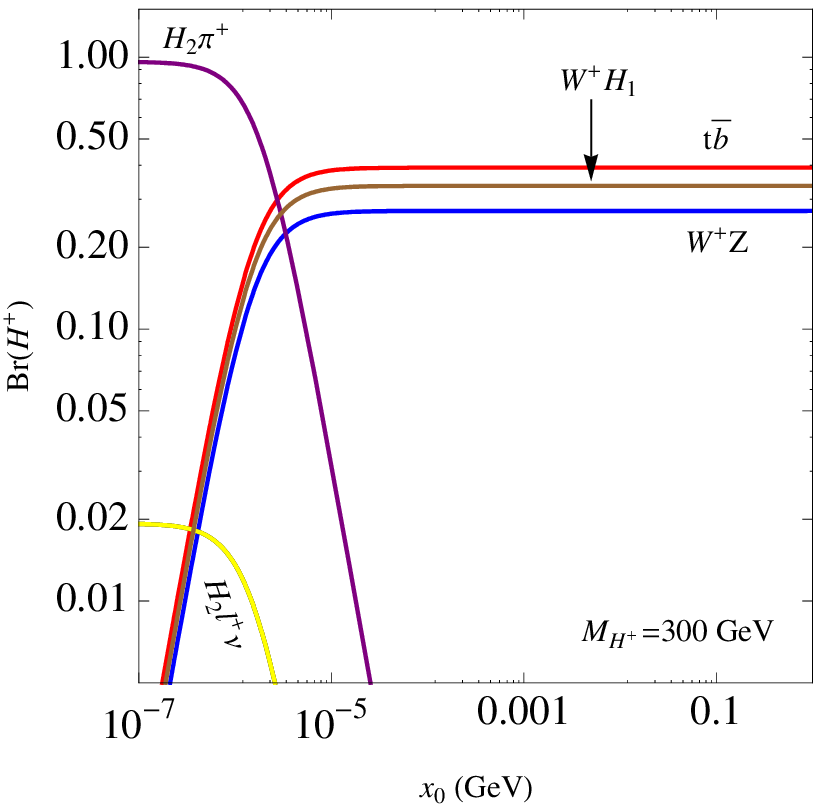}\hspace{0.03\textwidth}
\caption{Branching ratios for the singly charged Higgs
as a function of $x_0$ for $M_{H_2}=150$ GeV $-\Delta M$ in Eq.(24) (left panel) 
and $M_{H_2}=300$ GeV $-\Delta M$ (right panel). Here, we have taken  $M_{H_1}=120$ GeV.}
\label{fig:hpx0}
\end{figure}

\begin{figure}[h]
\includegraphics[scale=1,width=0.45\textwidth]{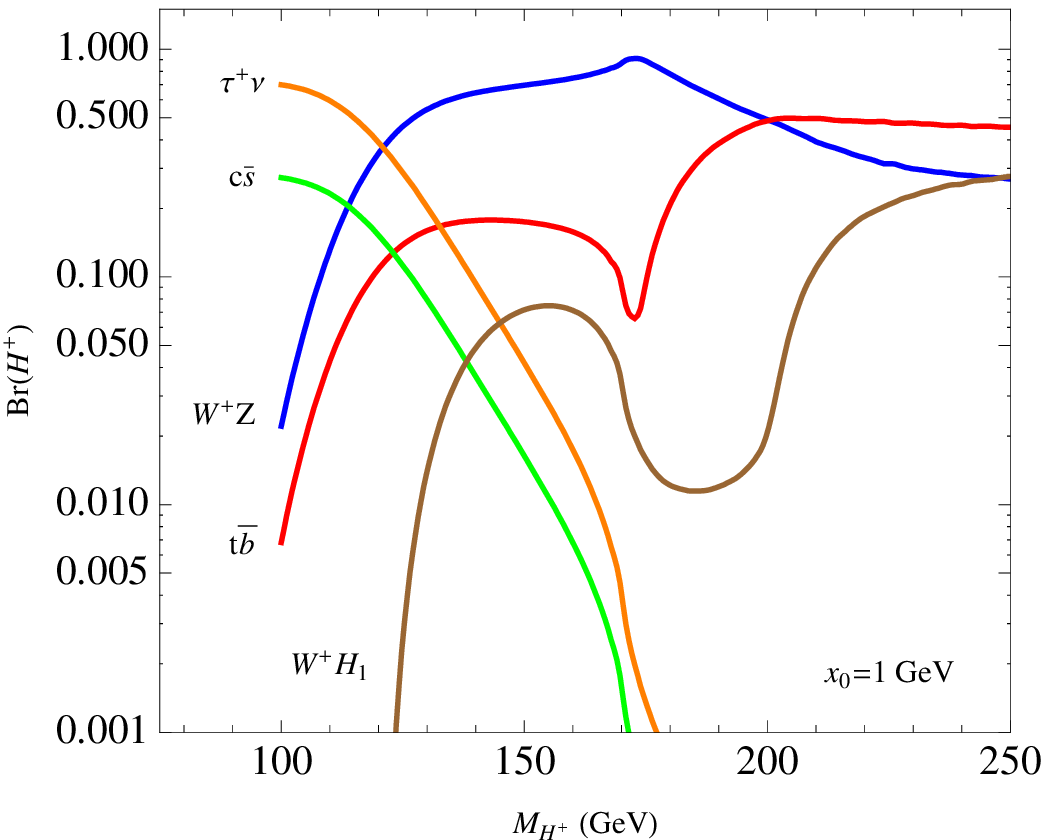}\hspace{0.03\textwidth}
\includegraphics[scale=1,width=0.45\textwidth]{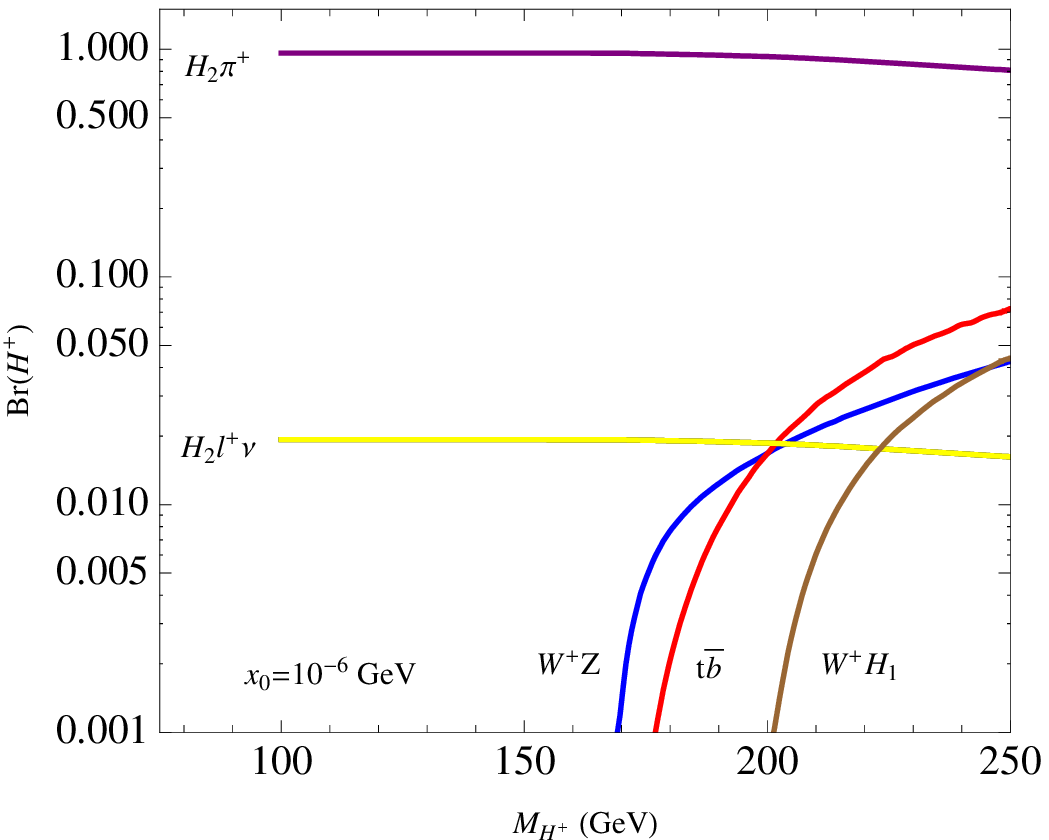}\hspace{0.03\textwidth}
\caption{Branching ratios for the singly charged Higgs as a function of $M_{H^+}$
when $x_0=1$ GeV (left panel) and $x_0=10^{-6}$ (right panel) using $M_{H_1}=120$ GeV.}
\label{hpx01GEV}
\end{figure}

As indicated earlier, the $H^{\pm}$ is never stable since $\Delta M > m_\pi$ in all cases. In the dark matter scenario, the
$H^\pm\to H_2\pi^\pm$ decay is the only two-body mode. The relative importance of this channel to other two-body modes depends
critically on the value of $x_0$ that governs the strength of the $H^\pm f{\bar f}$ Yukawa interaction via the mixing angle $\theta_+$.
In Fig. \ref{fig:hpx0}, we give the $H^\pm$ branching ratios as a function of $x_0$ for two illustrative values
of $M_{H^+}$. For  $M_{H^+}$ just below the $WZ$ threshold (left panel), Br($H^\pm\to H_2\pi^\pm$) dominates for $x_0\lesssim 10^{-4}$. For larger values of the triplet vev, the $W^\ast Z$ and $W Z^\ast$ channels are the largest, although the $t^\ast\bar{b}$ modes are also appreciable.
For heavier $H^\pm$ (right panel), the $t\bar{b}$, $WZ$ and $W H_1$ channels are leading when $x_0 \gtrsim 10^{-4}$.
The relative importance of the various final states for a given $x_0$ depends strongly on $M_{H^+}$, as illustrated in Fig. \ref{hpx01GEV}.
When the charged Higgs is light -- well below
the gauge-Higgs threshold --  the main decay channels for $x_0$ near the upper end of its allowed
range are $H^+ \to \tau^+ \nu$ and $H^+ \to c \bar{s}$ (see the left panel of Fig. \ref{hpx01GEV}). As $M_{H^+}$ is increased, the $WZ$, $WH_1$, and $t{\bar b}$ become dominant, with the relative importance of each depending on the specific range of $M_{H^+}$ under consideration. On the other hand, for very small $x_0$, the $H_2\pi^+$ final state dominates even for heavy $M_{H^+}$ (see the right panel of Fig. \ref{hpx01GEV}). These features of the $H^+$ decays can, in principle, be used both to distinguish the $\Sigma$SM from other scenarios as well as to determine the parameters $a_1$ and $x_0$. For the case of an unstable $H_2$, for example, $M_{H^+}^2\approx a_1 v_0^2/(4x_0)$ (see Eq.~(\ref{eq:mpls1})), while the branching ratios depend strongly on both $M_{H^+}$ and $x_0$. Thus, knowledge of both $M_{H^+}$ and the branching ratios could be used to identify the a range of values for these parameters.

\begin{figure}[h]
\includegraphics[scale=1,width=0.45\textwidth]{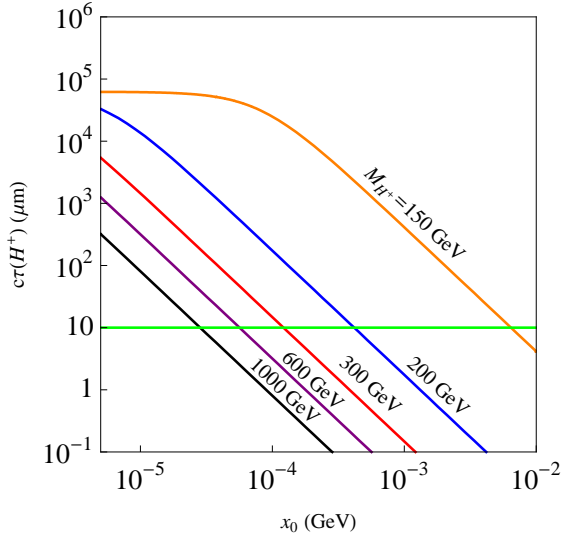}
\caption{Charged Higgs decay length as a function of $x_0$ for 
different values of $M_{H^+}$. The green line indicates the 
minimum needed for observation of a secondary vertex.}
\label{fig:hpmdecaylength}
\end{figure}

We emphasize that when the vev is very small, the charged Higgs is long-lived since
the total decay width is quite small. This feature can lead to the presence of a charged
track that can be used for identification. We illustrate this point in Fig. \ref{fig:hpmdecaylength}, where we show the decay length $c\tau_{H^+}$ as a function of $x_0$ for different values of $M_{H^+}$. For the decays above
the green line (horizontal line), one may observe a charged track associated with the $H^\pm$. It is important to
mention that the existence of the coupling $H^{+} W^{-} Z$ is due to the breaking of the custodial
symmetry once $\Sigma$ acquires a vev.  Recall that in a two Higgs doublet model this coupling
is absent. Therefore, one can use this decay in order to distinguish the model at future colliders.
%
\subsection{Triplet-Like Neutral CP-even Higgs Boson Decays}
The new extra neutral CP-even Higgs in this theory, $H_2$, is
triplet-like since the mixing in the neutral sector is typically
small due to the small allowed values of the triplet vev, $x_0 \lesssim 4 $ GeV.
At the same time, all the relevant couplings
of $H_2$ for the decays are suppressed by, $x_0$.  The total
decay width will be proportional to $x_0$, and
when $x_0 \to 0$, $H_2$ becomes stable and we recover the dark matter
scenario.
However, the branching ratios will be independent
of the triplet vev.
The specific branching ratios will differ from those for the SM-like Higgs due to the absence of a $\Sigma^0\Sigma^0 ZZ$ term in the Lagrangian and the dependence on $r$ in its coupling to $W^+W^-$. These features imply a change in the relative importance of the partial widths that depend on the $H_2 W^+ W^-$ coupling compared to the corresponding SM-like Higgs decays. Moreover, the $H_2$ branching ratios will depend strongly on the value of the quartic coupling $a_2$ due to its presence in $r$.

\begin{figure}[h]
\includegraphics[scale=1,height=7cm]{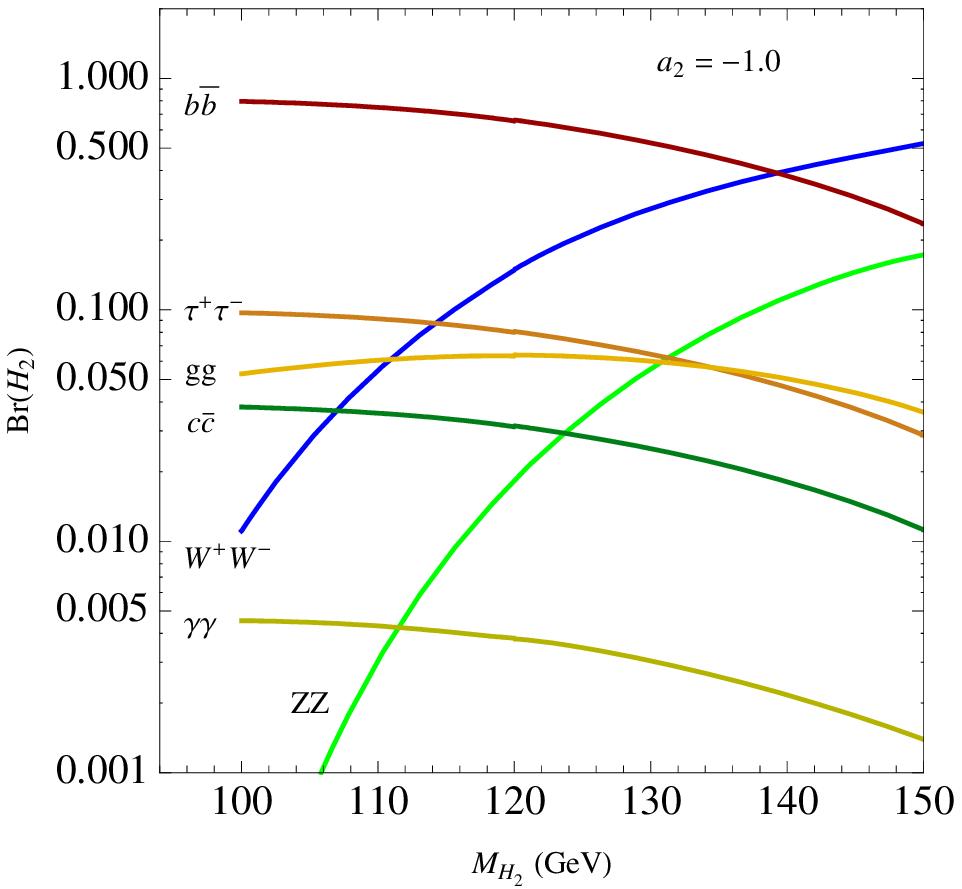}
\includegraphics[scale=1,height=7cm]{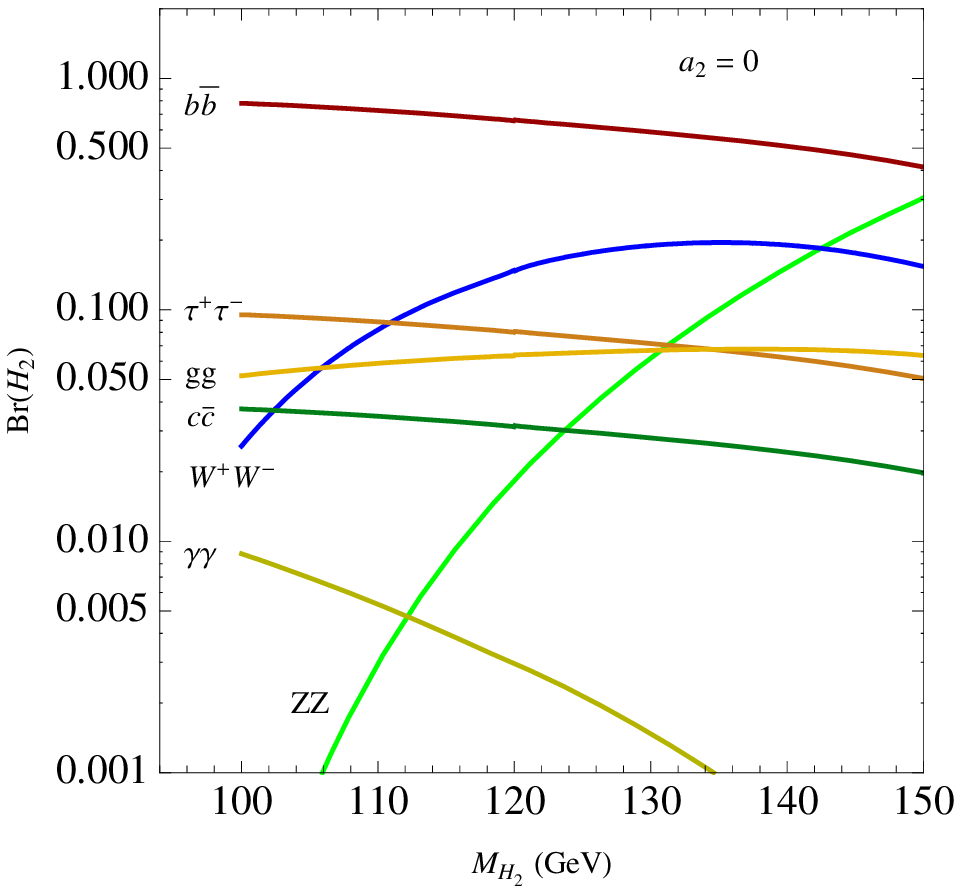}
\caption{Branching ratios for $H_2$ as a function of its mass
when $a_2=-1.0$ (left panel) and $a_2=0$ (right panel) using $M_{H_1}=120$ GeV.}
\label{nha2negative}
\end{figure}

\begin{figure}[h]
\includegraphics[scale=1,height=7cm]{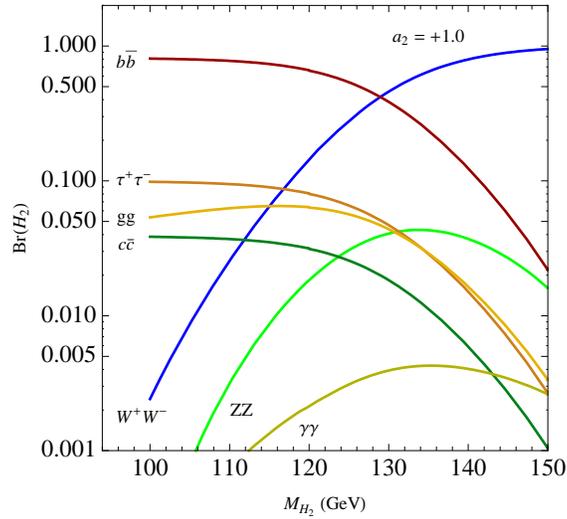}
\caption{Branching ratios for $H_2$ as a function of its mass when $a_2=+1.0$ using $M_{H_1}=120$ GeV.}
\label{nha2positive}
\end{figure}

Figures \ref{nha2negative}-\ref{nha2positive} illustrate the $H_2$ branching ratios as a function of
$M_{H_2}$ for different values of $a_2$. In each case, we see that when $H_2$ is light the most relevant decay channels are
$H_2 \to b \bar{b}, \tau^+ \tau^-, c \bar{c}, g g, W^\ast W$ and  $H_2 \to \gamma \gamma$.  The branching ratios 
for these channels are similar to those for the SM Higgs, except for the $W^\ast W$ and $\gamma\gamma$ channels. 
As discussed above, both depend on the $H_2 W^+ W^-$ coupling that does not require $\Sigma^0$-$h^0$ mixing to be 
non-vanishing. Consequently, the relative importance of these two branching ratios depends on the quartic coupling 
$a_2$. In particular, a relatively large, positive value for this parameter suppresses these branching ratios. In 
what follows, we will exploit the $\gamma\gamma$ channel in the strategy for discovery and identification of the $\Sigma$SM.
Once the massive gauge boson channels are open the relevant decays are $H_2 \to ZZ, H_1 H_1, W^+ W^-$
and $H_2 \to t \bar{t}$ and again the branching ratios are independent of $x_0$.
As Figs. \ref{nha2negative}-\ref{nha2positive} indicate, the branching ratios can
vary strongly with $a_2$ and can differ significantly from those for a pure SM Higgs.
For example, when $a_2=0$, the $ZZ$ branching ratio can be substantially larger
than that for a $WW$ final state, a situation that does not occur for the SM-like Higgs.
\begin{figure}[h]
\includegraphics[scale=1,width=7cm]{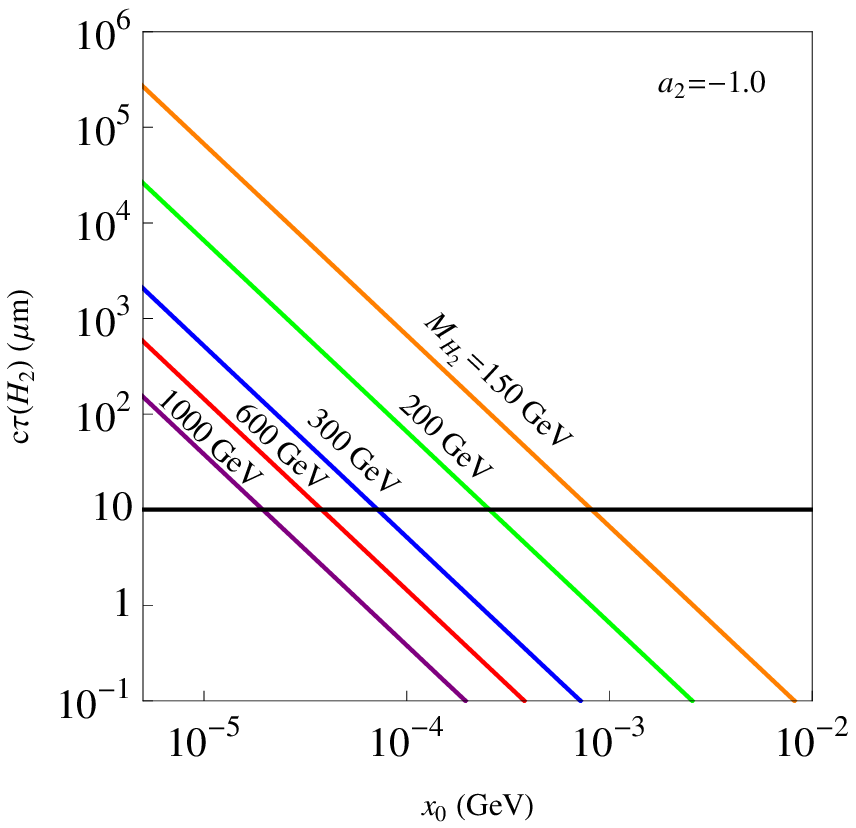}
\includegraphics[scale=1,width=7cm]{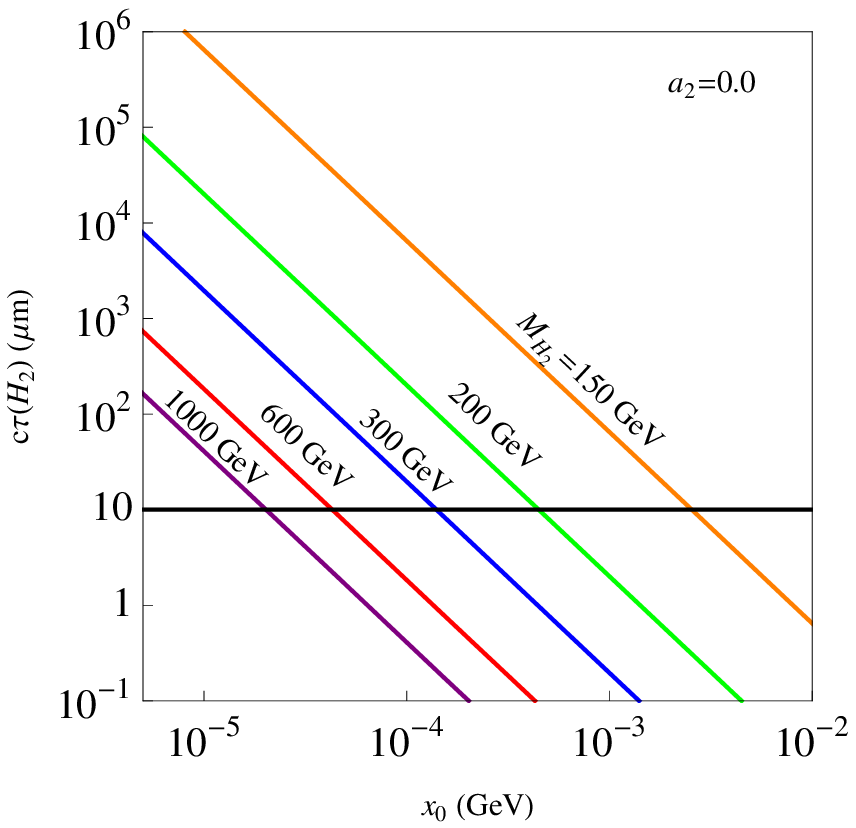}
\caption{Decay length of the heavy neutral Higgs versus
$x_0$ for $a_2=-1$ (left panel) and $a_2=0$ (right panel).}
\label{decaylength-H2-1}
\end{figure}
\begin{figure}[h]
\includegraphics[scale=1,width=7cm]{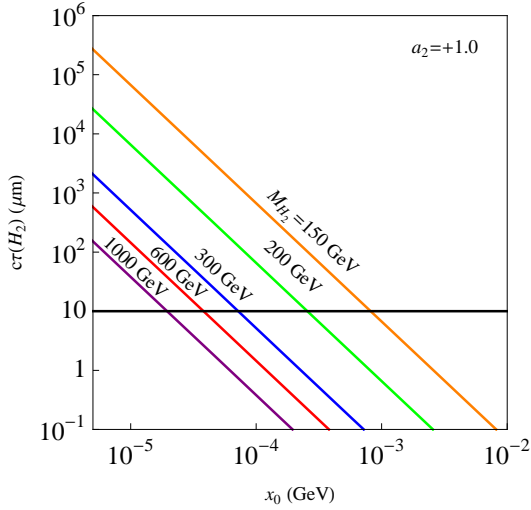}
\caption{Decay length of the heavy neutral Higgs versus $x_0$ when $a_2=1$.}
\label{decaylength-H2-2}
\end{figure}
In Figs.~\ref{decaylength-H2-1} and~\ref{decaylength-H2-2} we show the decay
length for the CP-even neutral triplet-like versus the triple
vacuum expectation value for different Higgs masses, where the
green line (horizontal line) corresponds to a decay length equal 
to 10 $\mu$m. Above this line one has a different scenarios
with a long-lived neutral Higgs and when $x_0 \to 0$ one recovers
the dark matter scenario.
\subsection{Heavy Higgs Scenario}
\begin{figure}[h]
\includegraphics[scale=1,width=7cm]{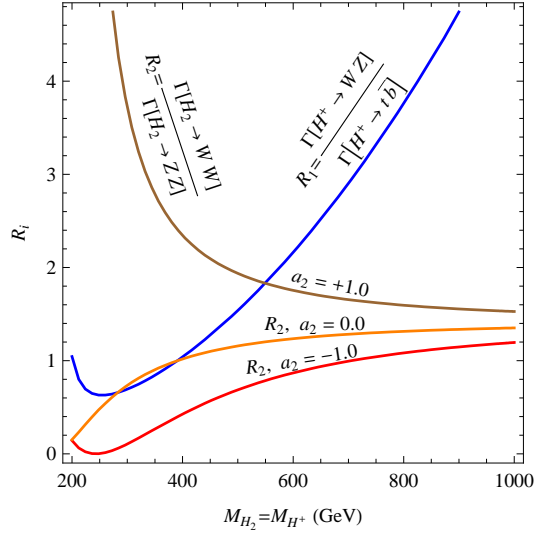}
\caption{Ratios between the various $H_2$ and $H^+$ decays when the scalars are heavy.}
\label{Ratios}
\end{figure}

When the mass of the Triplet-like Higgs is above the gauge boson
pair threshold one could in principle observe unique features of the $\Sigma$SM  at an $e^+e^-$ linear collider by
studying the ratios of different neutral and charged scalar decays. To illustrate this 
possibility, Fig.~\ref{Ratios} shows the predictions for the
ratios $R_1=\Gamma \left( H^+ \to W^+ Z \right) / \Gamma \left( H^+ \to t \bar{b}\right)$
and $R_2 = \Gamma \left( H_2 \to W W \right) / \Gamma \left( H_2 \to Z Z \right)$.
The ratio $R_1$ is always larger than one when the
Higgs mass is above 400 GeV, while
$\Gamma \left( H_2 \to W W \right) >  \Gamma \left( H_2 \to Z Z \right)$
only when the parameter $a_2$ is positive.
\section{Production Mechanisms at the LHC and Tevatron}
\label{sec:production}
\begin{figure}[ht]
\includegraphics[scale=1,width=8cm]{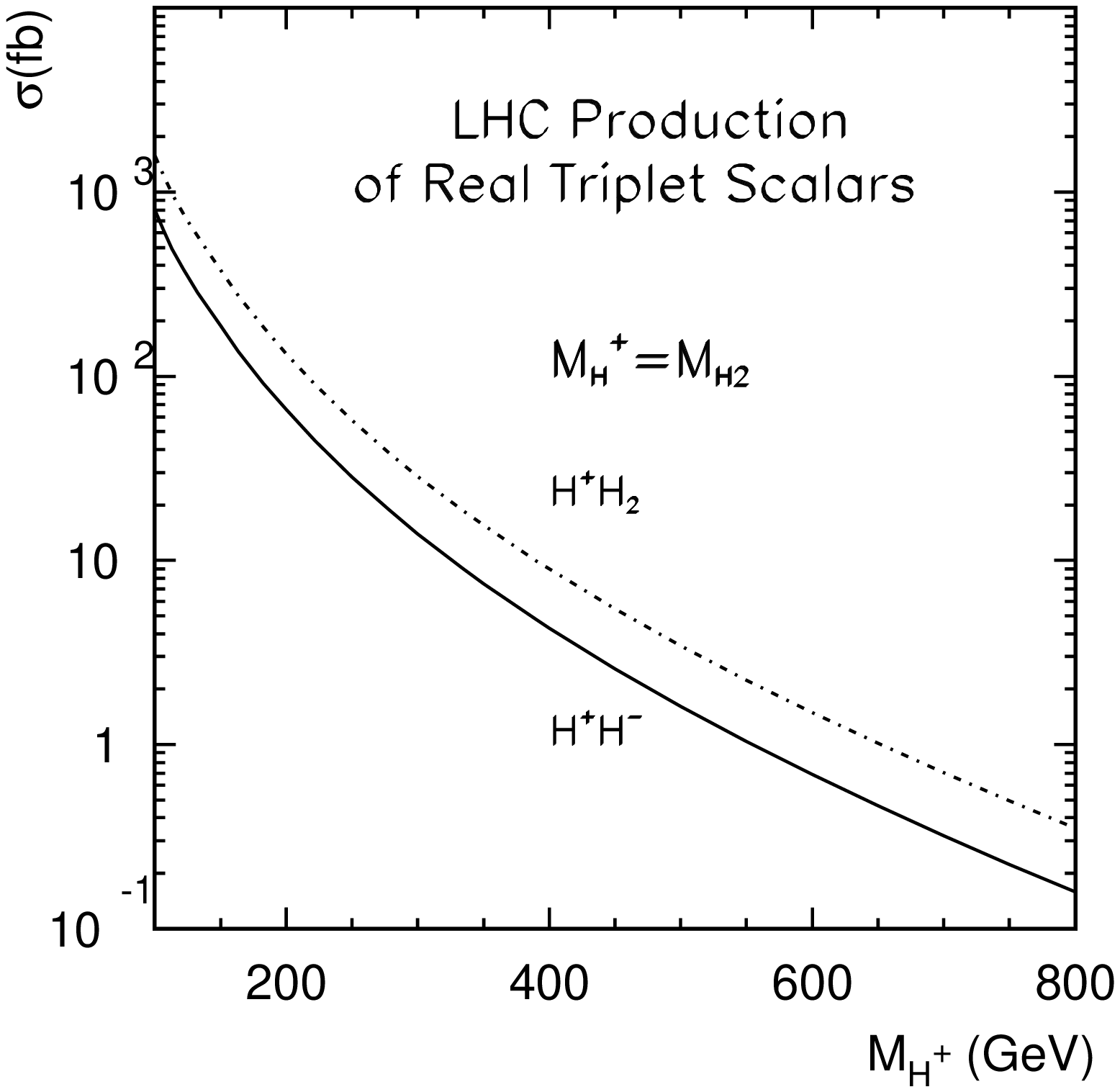}
\includegraphics[scale=1,width=8cm]{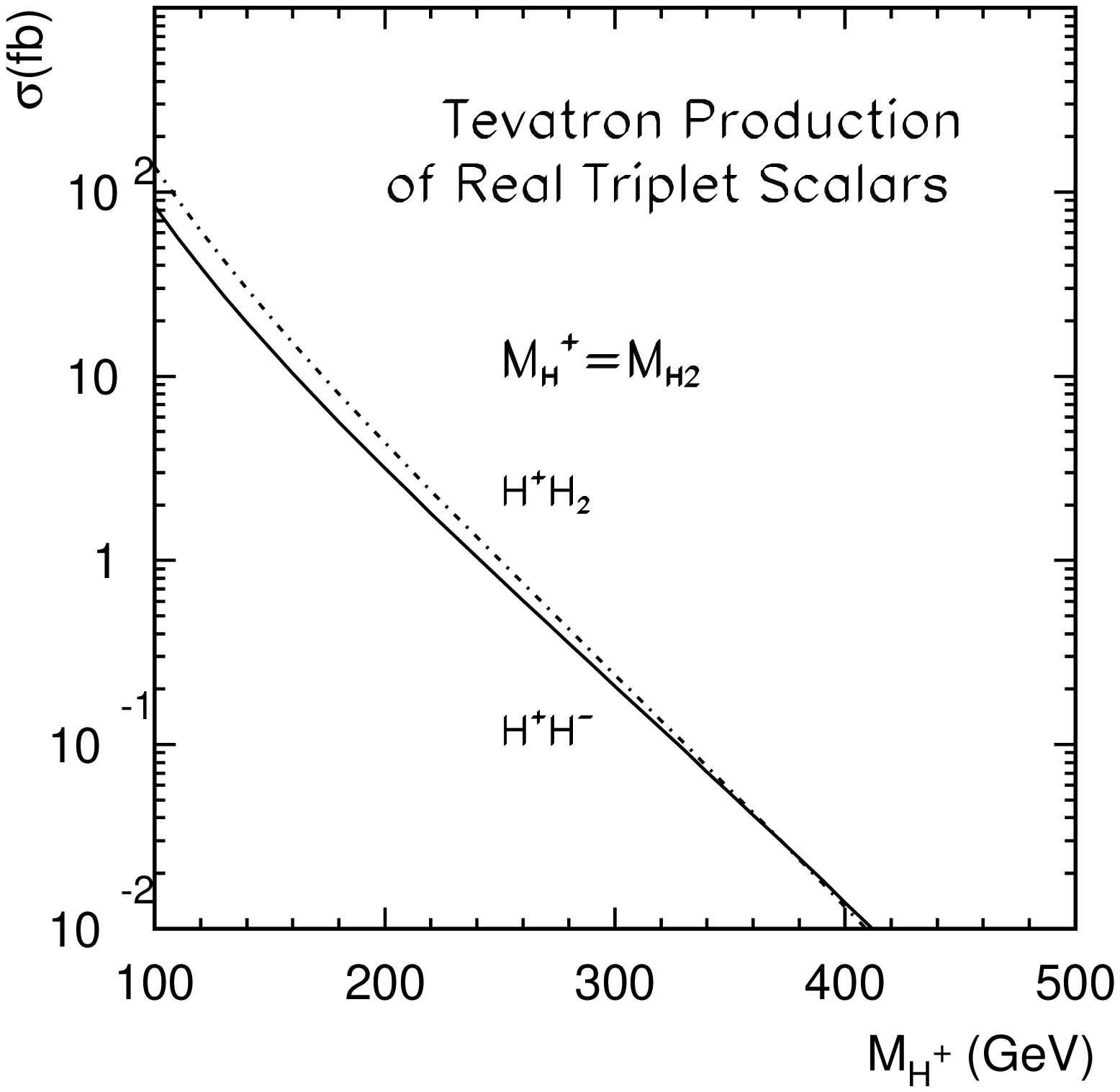}
\caption{Left panel: LHC production rate of $H^{\pm}H_2$  and $H^{+}H^{-}$.  Right panel: Tevatron production rate for
the same channels.}
\label{total1}
\end{figure}

In this section we study the production mechanisms for
$H^+$ and $H_2$ at the LHC. The leading production channels for
these scalars are the Drell-Yan (DY) pair production processes:
\begin{eqnarray}
q(p_1) \, + \, \bar{q}(p_2) \, &\rightarrow &
H^{+}(k_1)\, + \, H^{-}(k_2)\nonumber\\
q(p_1) \, + \, \bar{q}'(p_2) \, &\rightarrow &
H^{+}(k_1)\, + \, H_2(k_2)\nonumber
\end{eqnarray}
Here $p_i$ and $k_i$ are the momenta for the quarks and 
Higgses, respectively. In terms of the variable 
$y= \vec{p}_1\cdot \vec{k}_1/|p_1||k_1|$ in the parton center-of-mass frame
with energy $\sqrt{s}$, the parton level cross sections for
these processes are
\begin{eqnarray}
{d\sigma\over dy}(q \bar{q}\rightarrow H^{+}H^{-}) &=& \frac{3\pi \alpha^2 \beta_i^3 (1-y^2)}{4 N_c {s}}
\Big\{ e_q^2 +
\frac{ {s}}{({s}-M_Z^2)^2}
\frac{\cos 2\theta_W}{\tan^2 \theta_W}\nonumber\\
&&\qquad\times\Big[e_q g_V^q  ({s}-M_Z^2)\, + \, (g_V^{q2}+g_A^{q2}) {s}\  \frac{\cos 2\theta_W}{\tan^2\theta_W}
\Big]\Big\},
\enspace
\\
{d\sigma\over dy}(q \bar{q}'\rightarrow H^{\pm}H_2)& =& \frac{\pi \alpha^2 \beta_i^3(1-y^2)}{16 N_c \sin^4\theta_W}\frac{s}{(s-M^2_W)^2}\,,
\end{eqnarray}
where $\beta_i =\sqrt{(1-(m_i+m_j)^2/s)(1-(m_i-m_j)^2/s)}$ is the speed factor of $H_i H_j$ in the center-of-mass frame.
In the above equation $e_q$ is the electric charge of the quark and $N_c=3$, the number of colors. 
$g_V$ and $g_A$ are the vector and axial couplings, respectively.
In Fig.~\ref{total1} we plot the total  $H^+H^-$ and $H^\pm H_2$ production rate at the LHC and
Tevatron.

The QCD corrections to the process $H^{+}H^{-}$ and $H^\pm H_2$
are estimated from computation of $H^{++}H^{--}$ \cite{Tao} which are essentially equivalent. A next-to-leading (NLO) $K$-factor of order 1.25
at the LHC for Higgs mass range from 100 GeV to 1 TeV is expected \cite{qcd}.

\begin{figure}[ht]
\includegraphics[scale=1,width=8cm]{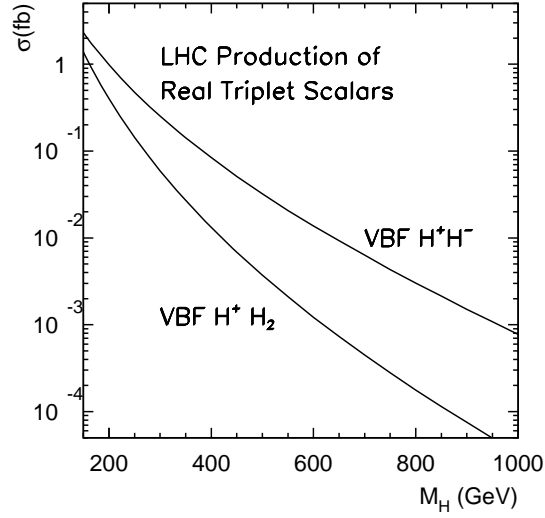}
\caption{ LHC production rate of VBF $jjH^{\pm}H_2$ and VBF $jjH^{+}H^{-}$ of real triplet model. }
\label{wbf1}
\end{figure}

The $H^\pm H_2$ and $H^+ H^-$ can also be produced via the weak vector-boson fusion (VBF) processes
\beq
\label{eq:vbf1}
pp  \rightarrow jj H^\pm H_2,\enspace jj H^+ H^-.
\eeq
The production rate is plotted in Fig. \ref{wbf1}. Since the Tevatron production
rate is very small and one will not be able to discover any event with $2 \text{fb}^{-1}$
integrated luminosity, we do not show the VBF rate for the Tevatron.
The VBF production rates are small compared with those for DY production,
but VBF offers the advantage of the production of
two very energetic forward/backward jets that helps
identifying events produced in this process.

In principle, a single triplet-like scalar can also be produced via the VBF processes
\beq
\label{eq:vbf1a}
pp  \rightarrow jj H_2, \enspace jj H^\pm \,.
\eeq
The presence of a three-body rather than a four-body final state implies less phase space suppression of the process in  Eq.~(\ref{eq:vbf1a}) compared to the processes in Eq.~(\ref{eq:vbf1}), so one might naively expect the single scalar production rate to be dominant.  However, the three-point couplings $W^+ W^- H_2$ and $W^\pm Z H^\mp$ are suppressed by a power of $x_0$ compared to the four-point couplings $W^+W^-H^+H^-$, $W^\pm ZH^\mp H_2$ and $ZZH^+H^-$ (see appendix A).  Therefore, the production rate of Eq.~(\ref{eq:vbf1a}) is in fact much smaller than that of Eq.~(\ref{eq:vbf1}). Similarly we argue that the associated
production  of single triplet scalars (Higgs-Strahlung),
\beq
pp\to W^\pm H_2, W^\pm H^\mp, Z H^\pm, Z H_2
\eeq
is also suppressed.  For this reason, we focus on  pair production of triplet-like  scalars
through DY or VBF processes.

The information provided by the plots summarizing the Higgs decay branching ratios
and decay lengths implies a variety of distinct search strategies for the additional
scalars in this model, for various values of the triplet vev, $x_0$, and the triplet-like scalar masses.
Here we outline three of the most promising avenues for the regime of light scalars, $M_H: 100-150~\text{GeV}$,
where the two-body decays to massive vector boson final states are kinematically
forbidden. In this low mass range, we will discuss three cases based on the $H_2$ behavior.
\begin{itemize}
\item[(i)]
$H_2$ is a matter candidate, with $x_0 = 0$: search for a monojet or monophoton and
one or two charged tracks in conjuction with missing transverse energy ($\cancel{E}_T$).
\item[(ii)]
$H_2\rightarrow \gamma\gamma$ for all allowed $x_0 \neq 0$: search for two photon events
in conjunction with a $\tau\nu$ final state or two $b$-jets.
\item[(iii)]
$H_2\to b\bar{b}$ for small vev $x_0 < 10^{-3}~\text{GeV}$: search for this mode 
in conjunction with the hadronic decays of the tau.
\end{itemize}
\subsection{Dark Matter Production and Search at LHC}
As discussed above, when $\Sigma \to - \Sigma$ and $x_0 = 0$,
$H_2$ is stable and $H_2$ and $H^\pm$ are degenerate at tree level.
The mass difference $\Delta M = M_{H^+}-M_{H_2}\simeq 166$ MeV arises
from radiative corrections. In this case the decays $H^\pm \to H_2 \pi^\pm$, $H^\pm \to H_2 \mu^\pm\nu_\mu$,
$H^\pm \to H_2 e^\pm\nu_e$ are the only allowed modes, with the total decay rate implying
a $c\tau_{H^\pm}=5.06$ cm \cite{Cirelli:2005uq}. The pions, electrons or muons
produced in the three-body decay are very soft and, thus, invisible
in the tracking system. In addition, the $H^\pm$ produced via DY or
VBF will not be highly boosted. Consequently, one should expect to see a charged track
of $\mathcal O(10~\text{cm})$ after which the $H^\pm$ becomes invisible. The charged Higgs
will mostly travel within the pixel detector~\cite{atlas}, so that the produced scalars and their decay products provide no
means for triggering.

For the trigger purpose, we consider  DY with mono-jet \cite{yanagida}, DY with mono-photon \cite{monophoton} or VBF production. In the monojet case, the Higgs pair will kick the jet, making it hard. One can then trigger on one hard jet
with large $\cancel{E}_T$. The monophoton trigger carries the same feature with less background. 
In the VBF case, the two forward/backward jet plus large $\cancel{E}_T$ will provide trigger selection rule.
\begin{figure}[ht]
\includegraphics[scale=1,width=8cm]{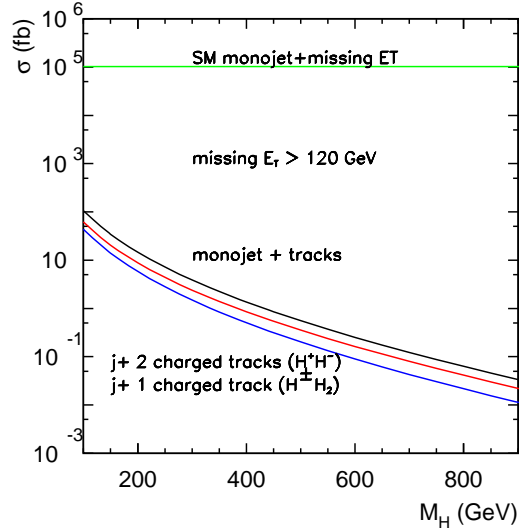}
\caption{LHC production of $j+H^\pm H_2$ and $jH^\pm H^\mp$.}
\label{monojet}
\end{figure}
Figure~\ref{monojet} shows the production rate of monojet+triplet scalar pair, $j\,H^\pm H_2$ and $jH^\pm H^\mp$.
The blue, red, and black curves give the rate for a singlet charged track ($H^\pm H_2$), two charged tracks ($H^+ H^-$), and their total, respectively. The trigger will be
\begin{itemize}
\item $p_T(j)>120$ GeV
\item $\cancel{E}_T> 120$ GeV
\end{itemize}
At the trigger level, one should expect a large background
from QCD $jZ$ with the $Z$ decaying invisibly and $jW$ with $W$ decaying
into soft leptons. The expected background rate is indicated by the green line.
To reduce this background, we impose a selection cut of $E^j_T > 120~\text{GeV}$,
and require at least one long lived charged particle with charged track length of greater than 5 cm,  then disappearing. With these additional criteria, the background is eliminated completely.

For the monophoton search, we employ the trigger as
\begin{itemize}
\item $p_T(\gamma)>50$ GeV
\item $\cancel{E}_T> 50$ GeV
\end{itemize}
and select the events with additional charged tracks.
\begin{figure}[ht]
\includegraphics[scale=1,width=8cm]{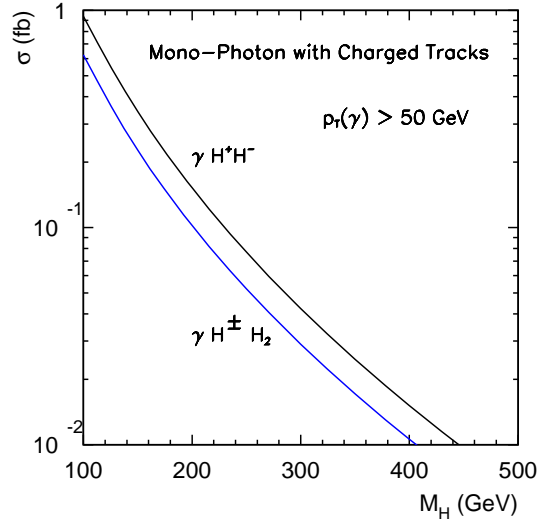}
\caption{LHC production of $\gamma H^\pm H_2$ and $\gamma H^\pm H^\mp$.}
\label{monophoton}
\end{figure}
The leading background at the trigger levels comes from the SM $\gamma Z$ which is 1201.7 fb.
Again, the event-selection requires at least one charged track with length $>5$ cm and the signal is
just event-counting.

\begin{figure}[ht]
\includegraphics[scale=1,width=8cm]{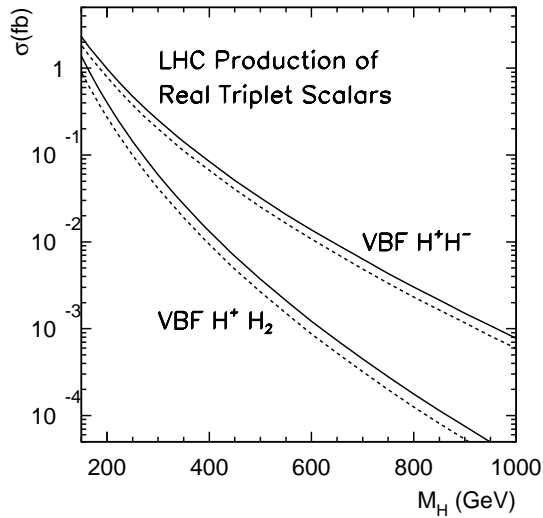}
\caption{VBF production of triplet Higgs pairs at the LHC with VBF selection cuts. Solid lines
are for basic cuts only and dashed lines are for VBF cuts.}
\label{wbf2}
\end{figure}

The VBF process carries a  unique feature of yielding two very energetic
forward/backward jets. The production rate of VBF triplet Higgs
pairs is shown in Fig. \ref{wbf1}. Again, we expect
large SM background at the trigger level associated with
QCD $jjZ$ events with $Z\to \nu\bar{\nu}$ and or $jjW$ with $W$ decay into soft leptons. To reduce this background we impose
the VBF selection cuts
\begin{itemize}
\item $p_T(j) > 50 $ GeV
\item $|\eta(j)|<5$
\item $\cancel{E}_T> 100$ GeV
\item $\eta(j_1) \cdot\eta(j_2) < 0$
\end{itemize}
The rates for VBF production of triplet scalars, after including these cuts, are shown in Fig. \ref{wbf2}.

In principle, we can also impose other cuts that identify the VBF features, such
as large dijet invariant mass ($M_{jj}$) or large absolute difference in rapidity, $|\Delta\eta_j|$.
There is no color exchange between the two jets and QCD jets will be
mostly in foward/backward region. Usually one can impose an additional
selection to reduce the QCD $jjX$ background. Initial simulations have
shown that the  VBF signal survival probability after central jet vetoing
is about 82\% at the LHC while the QCD $jjX$ processes
has only a 28\% survival probability using the central jet vetoing
procedure\cite{Rainwater:1999gg}. Fig.\ref{wbf2} has not included the central jet vetoing.
Similar to the monojet scenario, we will use the charged tracks
to select our signal events and the SM background will be eliminated
completely. Unfortunately, these events provide very little information that is useful for
measuring the triplet Higgs mass. One can only estimate the dark matter
mass through the observed production rate.
\subsection{$\gamma\gamma$-channel}
Photons do not appear in the tracking system but will deposit all their energy in the electromagnetic calorimeter.
Figures \ref{nha2negative}-\ref{nha2positive} suggests that $H_2\to\gamma\gamma$ may be a useful
channel to identify and reconstruct $H_2$, and it may even be used to
probe the parameter $a_2$.

To simulate detector effects on the photon energy-momentum measurements, we smear
the electromagnetic energy by a Gaussian distribution whose width is~\cite{atlas}
\begin{eqnarray}
{ \Delta E\over E} &=& {a_{cal} \over \sqrt{E/{\rm GeV}} } \oplus b_{cal}, \quad
a_{cal}=5\%,\  b_{cal}=0.55\%\,.
\label{ecal}
\end{eqnarray}
\begin{figure}[ht]
\includegraphics[scale=1,width=8cm]{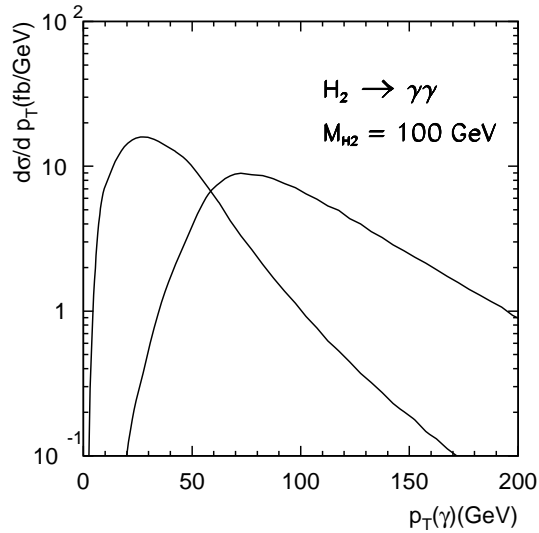}
\caption{$p_T(\gamma)$ distribution for $pp\rightarrow H^\pm H_2\rightarrow \tau^\pm \nu\gamma\gamma$ at the LHC.}
\label{total}
\end{figure}

The expected number of photon events depends strongly on $x_0$.  For relatively large $x_0$,
the decay mode $H^\pm\to\tau^\pm \nu$ is the dominant decay of $H^\pm$. Consequently, observation of a large number of
$\gamma\gamma\tau\nu$ events would indicate the large $x_0$ regime.  The $\tau^+\nu$ branching ratio is independent of $x_0$ for
large $x_0$ so a significant observation of these events would only indicate $x_0\gtrsim 10^{-3}$ GeV.  For much smaller $x_0$, the $H^\pm\to H_2\pi^\pm$
becomes leading, so we expect $H_2H_2+\pi^\pm$ and $H_2H_2+\pi^\pm\pi^\mp$ final states in this regime.
The $4\gamma$ final state will be extremely small due to the smallness of BR($H_2\to
\gamma\gamma$). Therefore, we recommend that one use the $\gamma\gamma b\bar{b}$ final states to identify the small vev regime.
See Fig. 16 for the $p_T$ distribution in the case of $pp\rightarrow H^\pm H_2\rightarrow \tau^\pm \nu\gamma\gamma$ at the LHC.

In the LHC enviroment, one expects that the diphoton events are usually easy to identify. 
In principle, the diphoton will also suffer from the jet faking photon events but 
to fake diphoton, the faking rate is of order $10^{-7}$ and this study will require
comprehensive detector simulation. We focus on the irreducible SM background as $\gamma\gamma+X$. 
We impose two hard photon selection cuts as (see Fig.~\ref{total})
\begin{itemize}
 \item $\text{\rm min}\{p_T(\gamma)\} > 25~ \text{GeV},\text{\rm max}\{p_T(\gamma)\} > 50~ \text{GeV} $
 \item $|\eta(\gamma)| < 2.8 $
 \item $\Delta R(\gamma\gamma),\, \Delta R(\gamma \ell) > 0.4$ ,
\end{itemize}
where $\Delta R=[\Delta\eta^2+\Delta \phi^2]^{1/2}$.

In the signal events, the diphoton decay steems from the triplet Higgs, 
hence one will expect a peak at diphoton invariant mass distribution around
the $M_H$.
\begin{figure}[ht]
\includegraphics[scale=1,width=8cm]{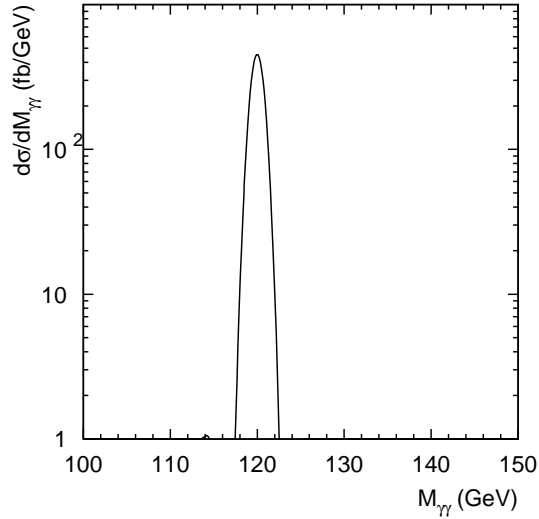}
\caption{$M_{\gamma\gamma}$ of $M_H=120$ GeV distribution for CMS resolution}
\label{fig:maa}
\end{figure}
Fig. \ref{fig:maa} shows that the diphoton invariant mass has much better resolution then dijet reconstruction.
Once identifying the peak in $M_{\gamma\gamma}$, we impose a selection cut
\beq
|M_{\gamma\gamma} - M_{H_2}| < 5~\text{GeV},
\eeq
where $M_{H_2}$ is the peak value.

To identify the $b{\bar b}$ final states associated with the diphoton, we require two $b$-tagged jets.
The leading SM background to this channel is $b\bar{b}(b)\gamma\gamma$. Consequently,  we
multiply the event number by the $b$-tagging efficiency of $(50\%)^2$ equal to 25\%. In addition, 
since the signals of $b$-jets are also from the triplet Higgs decays, one can construct dijet invariant 
mass $M_{bb}$ which is supposed to be equal to the $M_{\gamma\gamma}$ to confirm the 
triplet Higgs. To select the jet, we only propose the basic cuts as
\begin{itemize}
\item $p_T(b) > 15$ GeV,
\item $|\eta(b)|< 3.0$.
\end{itemize}

\begin{figure}[ht]
\includegraphics[scale=1,width=8cm]{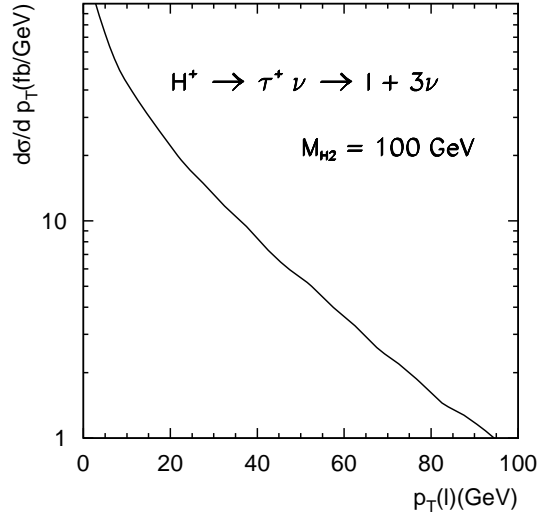}
\caption{Lepton transverse momentum, $p_T(\ell)$ from $\tau$ decay product of $H^+$.}
\label{taupt}
\end{figure}

The $\tau$ lepton has very different reconstruction from that of $\mu$ and $e$ in the detectors. The one-prong $\tau$ decay
BR is about 86\% with large missing $E_T$ and a single charged track. The charged track is generated by $\tau^+\rightarrow \pi^+\nu_\tau X$ ($X$ stands for neutral hadrons), $\tau^+\rightarrow e^+\nu_\tau\nu_e$ and $\tau^+\rightarrow \mu^+\nu_\tau\nu_\mu$. Fig. \ref{taupt} shows the $p_T$ of leptons from the $\tau$ 3-body decay for $M_H = 100$ GeV. The lepton $p_T$ here is only from the $\tau$ boost, and one expects the pions or leptons in this final state to be very soft.
The $\gamma\gamma\tau\nu$ final state also carries a unique feature as a diphoton with one single charged track
plus large missing $p_T$. However, QCD jet can fake the $\tau$-jet from $\tau$ hadronic decay. Consequently, we choose only
the $\tau$ leptonic decay which is 35\% of $\tau$ decay. Again since the leptons from $\tau$ decay are typically softer
than the leptons directly from $W$ decay, we will impose a cut as
\begin{itemize}
\item $p_T(l) > 5$ GeV, $p_T(l) < 40$ GeV
\item $|\eta(\ell)|< 2.8$
\item $\cancel{E}_T> 20$ GeV.
\end{itemize}
To confirm the $H^\pm\to \tau\nu$, one can use the $p_T$ of the track and the $\cancel{p}_T$ to
construct a transverse mass:
\beq
M_T = \sqrt{(E^\text{track}_T+\cancel{P}_T)^2-(\vec{p}^\text{\,track}_T+\vec{\cancel{p}}_T)^2}
\eeq
Using the edge of $M_T$, one can then reconstruct  $M_{H^\pm}$.

\begin{figure}[ht]
\includegraphics[scale=1,width=8cm]{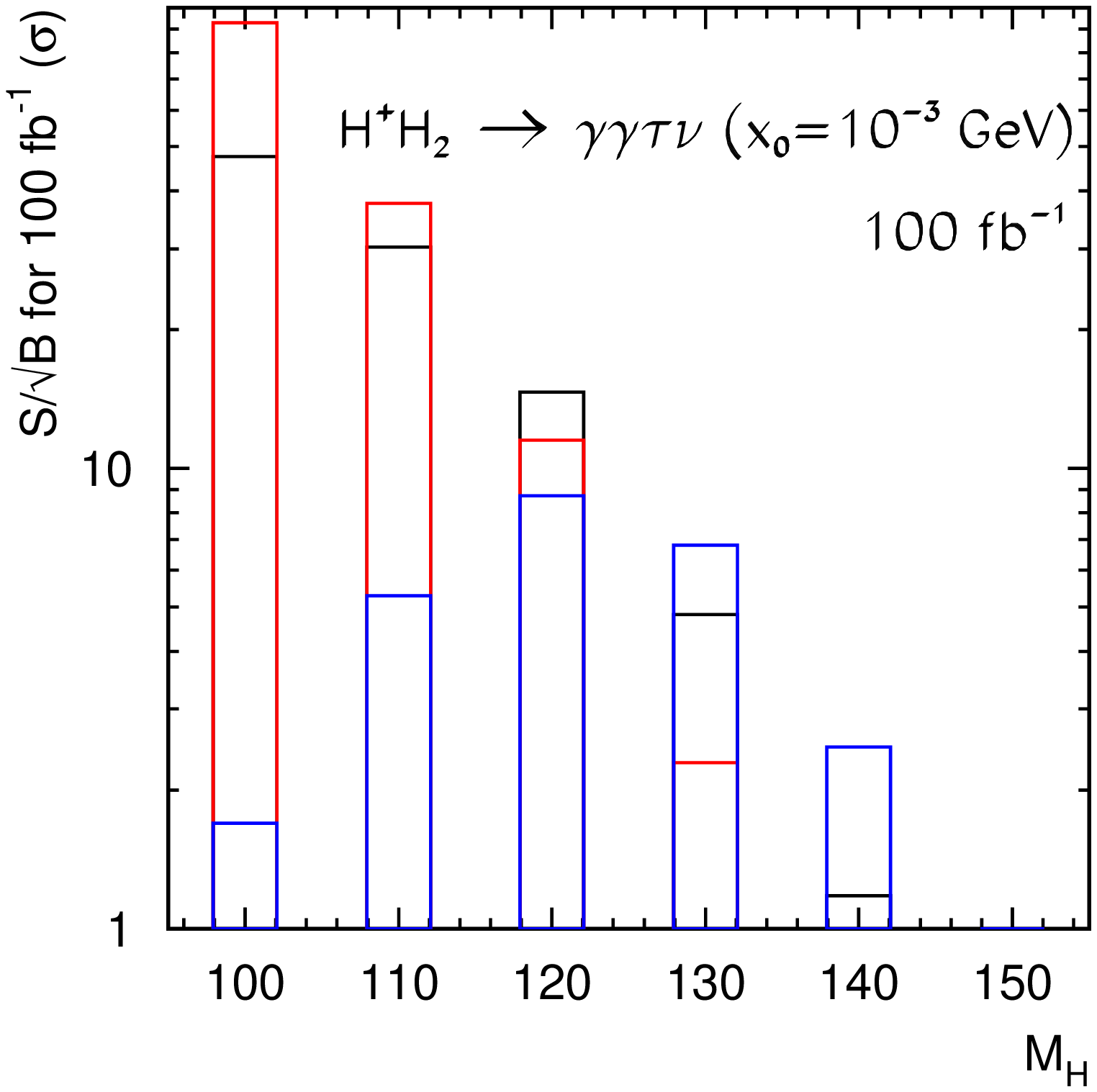}
\includegraphics[scale=1,width=8cm]{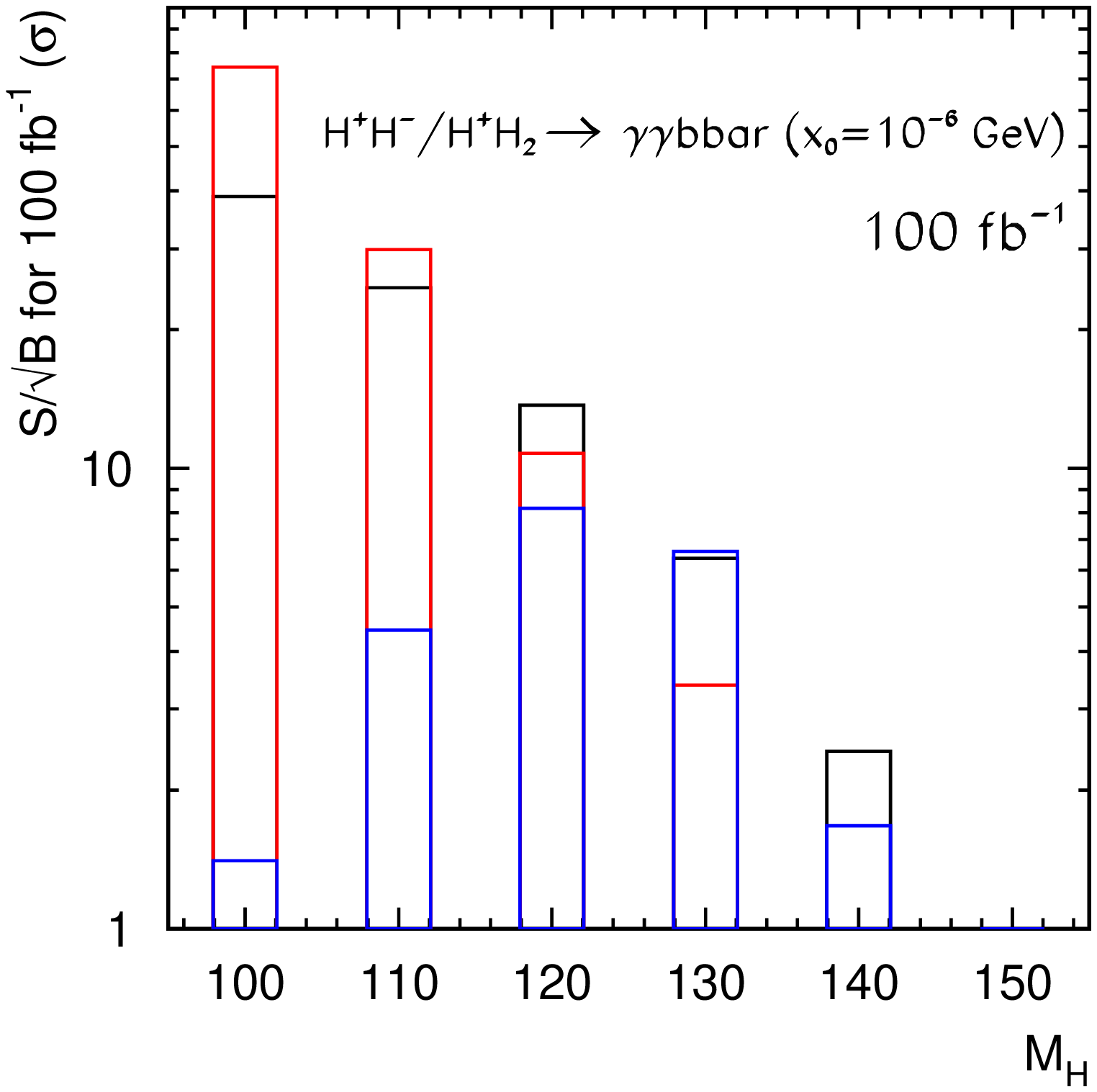}
\caption{$S/\sqrt{B}$ at 100 fb$^{-1}$ for $\gamma\gamma$, $x_0=10^{-3}~\text{GeV}$ and $x_0=10^{-6}~\text{GeV}$, where black, red and blue correspond to $a_2=-1,0,+1$ respectively.}
\label{aa}
\end{figure}

After imposing these cuts, the SM diphoton results are shown in Table I.
\begin{table}
\begin{tabular}{|c|c|c|c|}
\hline
$\sigma$ (fb) & Basic cuts & $M_{\gamma\gamma}$ cut & $p_T(l)$ cut \\
\hline
$b\bar{b}(b)\gamma\gamma$ & 11.59 & 0.78 & N/A\\
$W\gamma\gamma\to l \nu \gamma\gamma$ & 3.98  & 0.27 & 0.17 \\
$W\gamma\gamma\to \tau \nu \gamma\gamma\to l \gamma\gamma+{E_T}$ & 0.70  & 0.05 & 0.05\\
\hline
\end{tabular}
\caption{SM background to $\gamma\gamma$ events. For $bb\gamma\gamma$ final state, we require two $b$-tagged jets by assuming $b$-tagging efficiency of $50\%$.}
\end{table}
After the selection cuts, we plot the $S/\sqrt{B}$ in $\gamma\gamma\tau\nu$ and $\gamma\gamma b{\bar b}$, in Fig. \ref{aa} for 100 fb$^{-1}$ integrated
luminosity.

As emphasized above, the $H_2 W^+W^-$ coupling has a significant dependence on the parameter $a_2$. Due to the important $W^\pm$
one-loop contribution to the $H_2\gamma\gamma$ coupling, this $a_2$-dependence strongly affects the expected number of $\gamma\gamma X$ events. This
feature is also shown in Fig. \ref{aa}, where the black, red, and blue bars correspond to $a_2=-1$, 0, and $+1$, respectively. Given the number of measured events, one may infer a range on the value of $a_2$ using the value of $M_{H_2}$ that is obtained from the $M_{\gamma\gamma}$ reconstruction (see Fig. \ref{fig:maa}).

\subsection{$b\bar{b}$ final state}

As with the SM Higgs search, the $b\bar{b}$ is always the leading $H_2$ decay channel for the mass region
from 100 to 150 GeV before the $W^+W^-$ and $ZZ$ modes open up.  When $x_0$ is extremely small ($x_0 < 10^{-6}$ GeV), $H^\pm\to H_2\pi^\pm$ is the primary charged scalar decay, so we will expect the 4$b$ final state from $H^\pm H_2$ and $H^+H^-$ production to be dominant. This will encounter a huge SM QCD multijet background and will be impossible to be identified. In the $x_0 > 10^{-3}$ GeV region, for $H^\pm H_2$ production, $H^\pm \rightarrow \tau^\pm\nu$ is leading so that the  $b{\bar b} \tau^\pm \nu$  final state is the largest channel, while for $H^+H^-$ production, $\tau^+\tau^-\nu\bar{\nu}$ is leading. However, $H^+H^-\rightarrow \tau^+\tau^-\nu\bar{\nu}$ will be difficult to reconstruct. The presence of a $b\bar{b}$ in the final state helps in triggering, so we restrict our attention to $b{\bar b}\tau\nu$ channel in the regime where the $H^\pm\to\tau^\pm\nu$ branching fraction is significant.

We note that the production rate for this final state in the $\Sigma$SM is similar to its production in the 2HDM via the processes $q+q^\prime\to AH^\pm\to b{\bar b} \tau^\pm\nu$ and $q+q^\prime\to HH^\pm\to b{\bar b} \tau^\pm\nu$, where $A$ ($H$) is the neutral CP-odd (CP-even scalar) of that model\cite{bbtaunv}, as the $W^\pm H^\pm H_2$ and $W^\pm H^\pm A (H)$ couplings all have the same gauge coupling strength modulo scalar mixing angles. A study of the $b{\bar b}\tau\nu$ channel for the 2HDM was reported in Ref.~\cite{bbtaunv}, where it was shown that by observing the pion jet produced by the $\tau$ hadronic decay one could expect $S/\sqrt{B}\gtrsim 20$ for 100 $\mathrm{fb}^{-1}$ integrated luminosity at the LHC. As mentioned in the previous section, the study of the $\tau$ signature depends on the $\tau$ decay final state. In the study of Ref.~\cite{bbtaunv}, the authors used the feature that the $\tau^+$ produced in the decay $H^+\to\tau^+\nu$ arising from the Yukawa interaction is primarily left-handed, while the background $W^+$ bosons that decay to $\tau^+\nu$ have a primarily left-handed polarization and, thus, decay to primarily right-handed $\tau^+$ states. After boosting the angular distributions of the $\pi^+$ in the rest frame of the decaying $\tau^+$ along the direction of the $H^+$ or $W^+$ that produced it, one finds that the $p_T$ of the $\pi^+$ resulting from the $H^+$ decay chain is typically harder than that of the $\pi^+$ from the background $W^+$ decay chain. By imposing the cut
$p_T^\pi > 40$ GeV, the authors of Ref.~\cite{bbtaunv} suppress the $Wb{\bar b}$ background by a factor of four while reducing the signal event rate by $\sim 40\%$. We expect that a similar search strategy  using the pion jet for the $b{\bar b}\tau\nu$ final state in the $\Sigma$SM would yield a similar $S/\sqrt{B}$ and allow one to observe this channel effectively.

\begin{figure}[ht]
\includegraphics[scale=1,width=8cm]{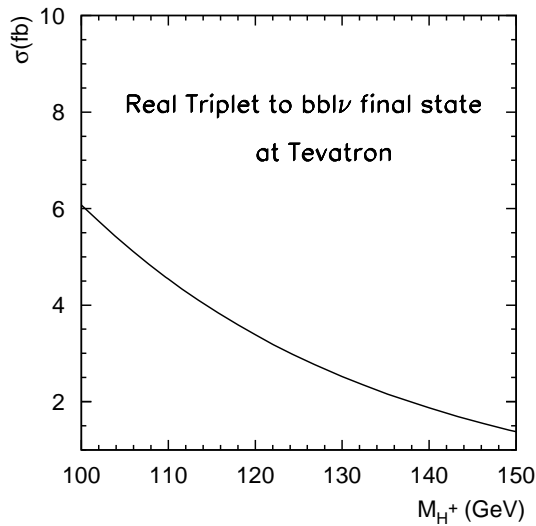}
\caption{Rate for $H^+H_2 \rightarrow \tau\nu b\bar{b}\rightarrow \ell b\bar{b} +\cancel{E}_T$ at the Tevatron using the SM Higgs search event selection criteria.}
\label{h1}
\end{figure}

For the $\tau$ three-body decay into lepton final states, the leptons are typically soft, and it is very challenging to search for such final states. However, it is still interesting that the $b\bar{b}\tau\nu$ final state is included in the Tevatron SM Higgs search via associated $WH$ production. Consequently, we have analyzed the possibility that the presence of the $\Sigma$SM could be observed through this Tevatron search. The conventional SM Higgs search criterion through associated production will cover part of the region for the $\tau\nu b\bar{b}$ final state associated with the $\tau$ leptonic decay. The leptons from $\tau$ leptonic decay are much softer compared with those from $W$-decays. One will expect a Jaccobian peak at $M_W/2$ for $p_T(\ell)$ from $W\rightarrow \ell \nu$ while the $p_T(\ell)$ from $\tau$ leptonic decay only comes from boost effects. The lepton $p_T$ cut for SM Higgs associated production search is
$$p_T(e) > 15~\text{GeV},~~p_T(\mu) > 10 ~\text{GeV}, ~~ |\eta(\ell)|< 2.8.$$
In addition to these cuts, we also include cuts for jets as
$$ p_T(j) > 25 ~\text{GeV}, |\eta(j)|< 3.0, |M_{jj}-M_H|<20 ~\text{GeV}. $$
The results are shown in Fig. \ref{h1}.
Our results indicate that, due to the hard lepton trigger, the $H^+H_2$ will only contribute to the SM Higgs signal at about the 10\% level.

\section{SUMMARY AND OUTLOOK}
The phenomenological aspects of a simple extension of the
Standard Model -- the $\Sigma$SM -- wherein the Higgs sector
is composed of the Standard Model $SU(2)_L$ doublet and a real
triplet have been investigated. Motivation for the $\Sigma$SM is both
theoretical and phenomenological: it may arise  as a low-energy remnant of
non-supersymmetric grand unified models that avoid rapid proton decay,
and if its neutral component has a vanishing vacuum expectation value
it provides a viable cold dark matter candidate. While the $\Sigma$SM has
been discussed extensively in the literature, its features relevant
to collider phenomenology have not been studied. Here, we have attempted to
provide such a study in order to determine how this scenario might be
discovered at the LHC, how it might be distinguished from other possible
extended Higgs scenarios, and how an analysis of collider observables
may provide information about the model parameters.

In general, we find that the $\Sigma$SM could be discovered at the LHC if the additional physical scalars -- the $H_2$ and $H^\pm$ -- are relatively light, with masses smaller than $\sim 150$ GeV, a regime in which two-body decays to massive vector bosons are kinematically forbidden. We find that there exist three distinct search strategies:
\begin{itemize}

\item[(1)] When the neutral component of the triplet is stable, the $H^\pm$ is long lived, yielding single or double charged track events. In DY production with a single initial state radiation for triggering we find that one could expect to see several hundred monojet plus track events with 100 fb$^{-1}$ of LHC running. In the mass range of interest, the $H_2$ would provide one element of a multi-component dark matter scenario.

\item[(2)] Whether or not the neutral component is stable, one could expect substantial deviations from the number of two-photon events associated with a SM Higgs boson. For a stable $H_2$, this effect arises from $H^\pm$ contributions to the $H_1\to\gamma\gamma$ amplitude. Depending on the value of the triplet mass and the quartic coupling $a_2$, the presence of these $H^\pm$ loops could lead to a doubling of the SM two-photon rate. When the $H_2$ is unstable, its two-photon decays could give rise to the $\gamma\gamma\tau\nu$ and $\gamma\gamma b{\bar b}$ events with a large $S/\sqrt{B} $ and 100 fb$^{-1}$ in $H^\pm H_2$ and $H^+ H^-$ DY production.

\item[(3)] When the triplet vev is very small, one may expect to identify $b{\bar b}\tau\nu$ events associated with a secondary vertex, allowing this final state to be distinguished from the very large SM backgrounds.

\end{itemize}

Assuming that one or more of these signatures is observed at the LHC, one could hope to identify ranges for the parameters of the model through a careful study of the scalar mass spectrum, branching ratios, and two-photon event rate.

When the new scalars associated with the $\Sigma$SM are above the two-body vector boson final state threshold, discovery and identification of the model at the LHC will be challenging at best. In this respect, a future $e^+e^-$ linear collider would in principle provide a more effective probe, not only  for discovery but also for identification of the model parameters as well. In either case, distinguishing the $\Sigma$SM from other models containing charged Higgs would likely require searching for unique features of those models, such as the CP-odd neutral scalar of the two Higgs doublet model or lepton-number violating final states (same sign dilepton pairs) in the seesaw triplet model.
\begin{acknowledgments}
We thank T. Han, F. Petriello, P. Langacker, C.P. Yuan and L.-T. Wang
for helpful discussions. This work was supported in part by the U.S. Department of
Energy contracts  DE-FG02-08ER41531 and DE-FG02-95ER40896 in part by the Wisconsin Alumni
Research Foundation. KW is supported by the World Premier International Research Center
Initiative (WPI Initiative), MEXT, Japan. We would like to thank T. Han
for providing private computing package hanlib.
\end{acknowledgments}

\appendix
\section{Vacuum Conditions}
\label{app:vac}
As stated in the text above, the model has four types of vacua depending on the choice of parameters.  Furthermore, the minimum corresponding to a phenomenologically viable vacuum may be accompanied by other local minima elsewhere.  A study of the theory requiring a scan across the parameter space must be restricted to regions for which the true vacuum yields the phenomenologically viable ones.  In what follows, we briefly present such conditions on the parameter set of the theory at tree level.  For convenience, we abbreviate the vacuum expectation values by the ordered pair $(\langle h^0\rangle,\,\langle\Sigma^0\rangle)$.  As in the text, we consider two cases:\\
\noindent \emph{Case (1): $a_1\neq0$}\\
The phenomenologically viable minimum occurs at $(h^0,\,\Sigma^0)=(v_0,\,x_0)$, with $v_0=246\text{ GeV}$.  We require that the extremization conditions, Eq. (\ref{min1}) and (\ref{min2}) are met.  We solve for $\mu^2$ and $M_\Sigma^2$,
\begin{gather}
\begin{aligned}\label{eq:MuAndM1}
\mu^2&=\lambda_0v_0^2-a_1x_0/2+a_2x_0^2/2\\
M_\Sigma^2&=b_4x_0^2+a_2v_0^2/2-a_1v_0^2/4x_0\,,
\end{aligned}
\end{gather}
and throughout the analysis, eliminate these parameters in favor of those appearing on the RHS of Eq. (\ref{eq:MuAndM1}).  We also require that the potential is concave upwards by requiring that both eigenvalues of the neutral mass matrix of Eq. (\ref{eq:massmtrx}) are positive.

We now consider possible minima that may accompany the physically viable one at $(v_0,\,x_0)$.  Each such candidate minima -- $(v,\,x)$, for example -- must satisfy its own extremization conditions analogous to Eq. (\ref{eq:MuAndM1}), which can be solved for $\mu^2$ and $M_\Sigma^2$.  In our example, we have
\begin{gather}
\begin{aligned}\label{eq:MuAndM2}
\mu^2&=\lambda_0v^2-a_1x/2+a_2x^2/2\\
M_\Sigma^2&=b_4x^2+a_2v^2/2-a_1v^2/4x\,.
\end{aligned}
\end{gather}
Now, we equate Eqs. (\ref{eq:MuAndM1}) with (\ref{eq:MuAndM2}) to formally eliminate $v$ and $x$ in favor of $v_0$ and $x_0$.
In order that the phenomenologically viable minimum, $(v_0,\,x_0)$, is the true vacuum of the theory, we demand that 
each candidate minima, $(v,x) \neq (v_0,x_0)$ is either (a) tachyonic: at least one of the two eigenvalues of the mass matrix, $\mathcal{M}^2$ evaluated at $(v,\,x)$ (see Eq. (\ref{eq:mmatrixfv}) below) is negative; or (b) a false vacuum: the potential at $(v,\,x)$ is shallower than the potential at $(v_0,\,x_0)$.  All conditions are expressed with $\mu^2$, $M_\Sigma^2$, $v$ and $x$ eliminated as described above.  We present in the table below conditions under which the phenomenologically viable minimum is the global minimum by considering three candidate minima.
\begin{equation*}
  \hspace{-2mm}\begin{array}{|c|cc|}
    \hline \substack{\displaystyle\text{Candidate}}&\text{(a)}\enspace\substack{\displaystyle\text{Tachyonic:}}
    &\text{(b)}\enspace\substack{\displaystyle\text{False Vacuum:}}\\\hline
    (0,\,0)&\begin{gathered}-(\lambda_0v_0^2+\textstyle\frac{1}{2}a_2x_0^2-\textstyle\frac{1}{2}a_1x_0)\leq0\text{   or}\\
                           -b_4x_0^2-\textstyle\frac{1}{2}(a_2v_0^2-\frac{1}{2}a_1v_0^2/x_0)\leq0\end{gathered}&
            \textstyle-\frac{1}{4}\lambda_0v_0^4-\frac{1}{4}b_4x_0^4+\frac{1}{8}(a_1v_0^2/x_0-2a_2v_0^2)x_0^2<0\\\hline
    (0,\,x)&\begin{gathered}\begin{aligned}-\lambda_0v_0^2+\textstyle\frac{1}{2}a_1(x_0-x)-\frac{1}{2}a_2(x_0^2-x^2)&\leq0\\
                           \text{or }b_4(3x^2-x_0^2)-\textstyle\frac{1}{4}(2a_2v_0^2-a_1v_0^2/x_0)&\leq0\,,\end{aligned}\\
                           \text{with }x^2=x_0^2+(2a_2v_0^2-a_1v_0^2)/4b_4\end{gathered}&
            \textstyle\lambda_0v_0^4/4-\frac{1}{64}(a_1v_0^2/x_0-2a_2v_0^2)^2/b_4>0\\\hline
    (v,\,x)&\begin{aligned}M_1(v,\,x)\leq0\\\text{or }M_2(v,\,x)\leq0\end{aligned}&\begin{aligned}\lambda_0(v_0^4-v^4)/4+b_4(x_0^4-x^4)/4\\ a_1(x_0-x)v^2/4+a_1v_0^2(x^2-x_0^2)/8x_0\\+a_2(x^2-x_0^2)(v^2-v_0^2)/4&>0\end{aligned}\\\hline
  \end{array}
\end{equation*}
where in the last case, $M_1(v,\,x)$ and $M_2(v,\,x)$ are eigenvalues of the mass matrix at $(v,\,x)$:
\begin{equation}\label{eq:mmatrixfv}
\mathcal{M}^2=\begin{pmatrix}\substack{\displaystyle\lambda_0(3v^2-v_0^2)-a_1(x-x_0)/2 + a_2(x^2-x_0^2)/2\\\phantom{.}}&(a_2x-a_1/2)v\\(a_2x-a_1/2)v&\substack{\phantom{.}\\
\displaystyle b_4(3x^2-x_0^2) + a_2(v^2-v_0^2)/2 + a_1v_0^2/4x_0}\end{pmatrix}
\end{equation}
and $x$ and $v$ are solutions to $\lambda_0v_0^2-a_1x_0/2+a_2x_0^2/2=\lambda_0v^2-a_1x/2+a_2x^2/2$ and $b_4x_0^2+a_2v_0^2/2-a_1v_0^2/4x_0=b_4x^2+a_2v^2/2-a_1v^2/4x$, as described above.

\noindent\emph{Case (2): $a_1=0$}\\
We follow the same procedure outlined above for case (1a).  The phenomenologically viable minimum occurs at $(h^0,\,\Sigma^0)=(v_0,\,0)$.  We require that the extremization condition, $-\mu^2+\lambda_0v_0^2+a_2x_0^2/2=0$ is met, and that the potential is concave upwards: $2\lambda_0v_0^2>0$ (implying $\lambda_0>0$) and $a_2v_0^2/2-M_\Sigma^2>0$.  The value of the potential at this minimum is $V(v_0,\,0)=-\lambda_0 v_0^2/4$.  The table below summarizes the conditions underwhich the phenomenologically viable minimum is the global minimum.
\begin{equation*}
  \begin{array}{|c|cc|}
    \hline \substack{\displaystyle\text{Candidate}}&\text{(a)}\enspace\substack{\displaystyle\text{Tachyonic:}}&\text{(b)}\enspace
    \substack{\displaystyle\text{False vacuum:}}\\\hline
    (0,\,0)&-\lambda_0v_0^2<0\enspace\text{ or }-M_\Sigma^2<0&\\\hline
    (0,\,x)&\begin{gathered}\begin{aligned}-\lambda_0v_0^2+a_2M_\Sigma^2/(2b_4)&\leq0\\\text{or }2M_\Sigma^2&\leq0\end{aligned}\\
    \text{where }M_\Sigma^2=b_4x_0^2+a_2v_0^2/2 \end{gathered}&\lambda_0b_4v_0^4>(b_4x_0^2+a_2v_0^2/2)^2\\\hline
(v,\,x)&\text{see below}&\\\hline
\end{array}
\end{equation*}
Notice that since $\lambda_0>0$, the condition for candidate minimum $(0,\,0)$ to be tachyonic is already satisfied.
\begin{list}{\arabic{Lcount}.}{A global minimum at $(v,\,x)$ can be avoided by the following conditions:}
\usecounter{Lcount}
\item Either $v$ or $x$ in terms of $v_0$ and $x_0$ is complex:
\begin{gather}\label{eq:VandX}
    v=\pm\left(\frac{\lambda_0v_0^2-a_2M_\Sigma^2/2b_4}{\lambda_0-a_2^2/4b_4}\right)^{1/2}\,,\enspace
    x=\pm\left(\frac{M_\Sigma^2}{b_4}-\frac{a_2}{2b_4}\frac{\lambda_0v_0^2-a_2M_\Sigma^2/2b_4}{\lambda_0-a_2/4b_4}\right)^{1/2}\,,
\end{gather}
where $M_\Sigma^2=b_4x_0^2+a_2v_0^2/2$.
\item Otherwise, the minimum at ($v$, $x$) contains a tachyonic mode: at least one of the two eigenvalues,
\begin{gather}
    M_\pm^2=(\lambda_0v^2+b_4x^2)\pm\left((\lambda_0v^2+b_4x^2)^2-a_2^2v^2x^2\right)^{1/2}\,,
\end{gather}
of the mass matrix is negative.
\item Else, the potential at $(v, x)$ is shallower than at $(v_0,\,0)$:
\begin{gather}
0<\frac{\lambda_0}{4}(v^2-v_0^2)^2+\frac{b_4}{4}x^4-\frac{1}{2}M_\Sigma^2x^2+\frac{1}{4}v^2a_2x^2\,,
\end{gather}
where $x$ and $v$ are given by Eq. (\ref{eq:VandX}).

\end{list}

\section{Formulae for Partial Widths of $H_1$, $H^+$ and $H_2$}
\begin{flushleft}
\underline{Triplet-Like Neutral Scalars $H_2$}:
\end{flushleft}
\begin{align}
\Gamma(H_2\rightarrow V^*\,V)&=\frac{3 G_F^2
|g_{H_2VV}|^2M_V^4}{16\pi^3M_{H_2}}\delta'_V\int_0^{M_V^2}dM_*^2\,
\frac{\beta_V(M_{H_2}^2\beta_V^2+12
M_V^2M_*^2)}{(M_*^2-M_V^2)^2+M_V^2\Gamma_V^2}\,,
 \intertext{where $\delta'_W=1$, $\delta_Z=\frac{7}{12}-\frac{10}
{9}\sin^2\theta_W+\frac{40}{9}\sin^4\theta_W$,
 and
 $\beta_V^2=\displaystyle\left(1-\frac{(M_V+M_*)^2}{M_{H_2}}\right)
\left(1-\frac{(M_V-M_*)^2}{M_{H_2}}\right)$}
\Gamma(H_2\rightarrow
H_1^*H_1)&=\frac{3G_F^2}{32\pi^3}\frac{M_Z^4}{M_H}\cos^2\theta_0M_b
\int_0^{1-r_{H_1}^2}dx_2\int_{1-x_2-r^2_{H_1}}^{1-r^2_{H_1}/(1-
x_2)}dx_1\,\frac{x_1+x_2-1+r^2_{H_1}}{(1-x_1-
x_2)^2+r_{H_1}^2\Gamma_{H_1}^2/M_{H_2}}
\\
\Gamma(H_2\rightarrow f\bar{f})&=\frac{N_C}{16\pi}|g_{H_2f\bar{f}}|
^2M_{H_2}(1-4r_f^2)^{3/2}
\\ & \nonumber \\
\label{HggDecay}\Gamma(H_2\rightarrow gg)&=\frac{\alpha_s
g_2^2}{128\pi^3}\frac{M_{H_2}^3\sin^2\theta_0}{M_W^2}
[4r_t^2(1+(1-4r_t^2)f(4r_t^2))]
\end{align}
\begin{multline}
\Gamma (H_2 \to \gamma \gamma) = \frac{\alpha^2 g^2_2}{1024
 \pi^3}\frac{M_{H_2}^3}{M_W^2}\bigg|  \frac{M_W}
{M_{H_2}^2}\frac{g_{H_2 H^+ H^-}}{g_2} F_0(4 r_{H^+}^2)\\ + \frac{8}
{3\sqrt{2}}\frac{M_W}{g_2\,M_t}g_{H_2t\bar{t}} F_{1/2}(4r_t^2)
+  \frac{g_{H_2 W^-W^+}}{M_W} F_1(4r_W^2)  \bigg|^2,
\end{multline}
where the loop functions are
\begin{equation}
\begin{aligned}
F_0(x)&=x(1-x\,f(x))\\
F_{1/2}(x)&=-2x\big(1+(1-x)f(x)\big)\\
F_1(x)&=2+3x+3x(2-x)f(x)\\
\end{aligned}
\enspace
\text{with}\enspace f(x)=
\left\lbrace\begin{aligned}&[\sin^{-1}\big(\sqrt{1/x}\big)]^2,&x\geq1
\\ &\textstyle\frac{-1}{4}\big[\ln(\frac{1+\sqrt{1-x}}{1-\sqrt{1-x}})-i
\pi\big]^2&x<1,\end{aligned}\right.
\end{equation}
and $\Gamma_i$ is the total width of particle $i$. $N_C=3$ for quarks,
$N_C=1$ for leptons, and $r_i=M_i/M_H$ is the ratio of
masses of particle $i$ to decaying scalar boson.  See Ref.~
\cite{EWPO} for the expressions
of the decay rates.
\section{Feynman Rules}
\begin{center}
\begin{table}[h]
\begin{tabular}[t]{|c||c|c|}
\hline Interaction & Feynman Rule\footnote{Feynman rules are given such that all momenta flow into the vertex.}\\
\hline
$H_1f\bar{f}$&$i(M_f/v_0)\cos\theta_0$\\
$H_2f\bar{f}$&$-i(M_f/v_0)\sin\theta_0$\\
$H^+ \bar{u} d$ & $-i \frac{\sqrt{2}}{v_0} \sin \theta_{+} \ \bar{u} \left( - m_u \ V_{CKM} \ P_L \ + \ V_{CKM} \ m_d  \ P_R \right)
d \ H^{+}$ \\
\hline \hline
$H_1H_1H_1$ &
$-i\big(3x_0a_2c_0^2s_0+3v_0a_2c_0s_0^2+\frac{3}
{2}a_1c_0^2s_0+6b_4x_0s_0^3+6\lambda_0v_0c_0^3\big)$\\
$H_2H_1H_1$&$\frac{1}{2}a_1c_0^3-a_1c_0s_0^2+2a_2v_0c_0^2s_0-
a_2v_0s_0
^
3
+a_2x_0c_0^3-2a_2x_0c_0s_0^2+6b_4x_0c_0s_0^2-6\lambda_0v_0c_0^2s_0\big)
$\\
$H_1H_2H_2$&$-i\big(\frac{1}{2}a_1s_0^3-
a_1c_0
^
2s_0
-2a_2v_0c_0s_0
^
2
+
a_2v_0c_0
^3+a_2x_0s_0^3-2a_2x_0c_0^2s_0+6b_4x_0c_0^2s_0+6\lambda_0v_0c_0s_0^2$\\
$H_2H_2H_2$&$-i\big(3a_2x_0c_0s_0^2-3a_2v_0c_0^2s_0+\frac{3}
{2}a_1c_0s_0^2+6b_4x_0c_0^3-6\lambda_0v_0s_0^3\big)$\\
$H^+H^-H_1$&$-i\big(a_1c_+s_+c_0-\frac{1}{2}a_1s_+^2s_0+a_2v_0c_
+^2c_0+a_2x_0s_+^2s_0+2b_4x_0c_+^2s_0+2\lambda_0v_0s_+^2c_0\big)$\\
$H^+H^-H_2$&$-i\big(-a_1c_+s_+s_0-\frac{1}{2}a_1s_+^2c_0-a_2v_0c_
+^2s_0+a_2x_0s_+^2c_0+2b_4x_0c_+^2c_0-2\lambda_0v_0s_+^2s_0\big)$
\\\hline\hline
$ZZH_1$&$(2iM_Z^2/v_0)c_0g^{\mu\nu}$\\
$ZZH_2$&$-\frac{2iM_Z^2}{v_0}s_0g^{\mu\nu}$\\
$ZW^\pm H^\mp$&$ig_2\big(-g_2x_0c_+c_w+\frac{1}{2}g_1v_0s_+s_w\big)g^{\mu
\nu}$\\
$W^+W^-H_1$&$ig^2_2\big(\frac{1}{2}v_0c_0+2x_0s_0\big)g^{\mu\nu}$\\
$W^+W^-H_2$&$ig^2_2\big(-\frac{1}{2}v_0s_0+2x_0c_0\big)g^{\mu\nu}$\\
$\gamma H^+H^-$&$ie\,\big(p'-p\big)^\mu$\\
$ZH^+H^-$&$i\big(g_2c_w-\frac{M_Z}{v_0}s_+^2)\big(p'-p\big)^\mu$\\
$W^\pm H_1H^\mp$&$\pm ig_2\big(\frac{1}{2}s_+c_0+
c_+s_0\big)\big(p'-p\big)^\mu$\\
$W^\pm H_2H^\mp$&$\mp ig_2\big(\frac{1}{2}s_+s_0-c_+c_0\big)\big(p'-p
\big)^\mu$
\\\hline\hline
$W^+W^-H^+H^-$&$-\frac{i}{2}g_2^2 c_+^2$\\
$W^\pm Z H^\mp H_2$&$-i(g_2^2 c_+c_0c_w+\frac{1}{2}eg_1s_+s_0)$\\
$ZZH^+H^-$&$i(2e^2c_+^2+\frac{1}{8 M_Z^2}s_+^2v_0^2(g_2^2-g_1^2))$\\
\hline
\end{tabular}
\label{tab:feynman}
\end{table}
\end{center}


\end{document}